\documentclass[acmsmall,authorversion]{acmart}

\AtBeginDocument{%
  \providecommand\BibTeX{{%
    \normalfont B\kern-0.5em{\scshape i\kern-0.25em b}\kern-0.8em\TeX}}}

\setcopyright{acmlicensed}
\acmJournal{PACMHCI}
\acmYear{2023} \acmVolume{7} \acmNumber{CSCW2} \acmArticle{280} \acmMonth{10} \acmPrice{15.00}\acmDOI{10.1145/3610071}

\usepackage{booktabs,caption}
\usepackage[flushleft]{threeparttable}
\usepackage{wrapfig}
\usepackage{amsmath}
\usepackage{multirow}

\begin{document}

\title[Newcomer Homepage]{Increasing Participation in Peer Production Communities with the Newcomer Homepage}

\author{Morten Warncke-Wang}
\email{mwang@wikimedia.org}
\orcid{0000-0002-0621-4048}
\author{Rita Ho}
\email{rho@wikimedia.org}
\orcid{0009-0004-4677-8851}
\author{Marshall Miller}
\email{mmiller@wikimedia.org}
\orcid{0009-0001-8108-0182}
\author{Isaac Johnson}
\email{isaac@wikimedia.org}
\orcid{0000-0002-8869-3010}
\affiliation{
  \institution{Wikimedia Foundation}
  \country{USA}
}

\renewcommand{\shortauthors}{Morten Warncke-Wang et al.}

\begin{abstract}
    For peer production communities to be sustainable, they must attract and retain new contributors. Studies have identified social and technical barriers to entry and discovered some potential solutions, but these solutions have typically focused on a single highly successful community, the English Wikipedia, been tested in isolation, and rarely evaluated through controlled experiments. We propose the Newcomer Homepage, a central place where newcomers can learn how peer production works and find opportunities to contribute, as a solution for attracting and retaining newcomers. The homepage was built upon existing research and designed in collaboration with partner communities. Through a large-scale controlled experiment spanning 27 non-English Wikipedia wikis, we evaluate the homepage and find modest gains, and that having a positive effect on the newcomer experience depends on the newcomer's context. We discuss how this impacts interventions that aim to improve the newcomer experience in peer production communities.
\end{abstract}



\maketitle

\section{Introduction}
Sustainable peer production communities~\cite{benkler2002coase,benkler2015peer} must attract and retain newcomers. Wikipedia is one such community, whose mission is to collect and develop free educational content and disseminate it effectively and globally. Maintaining such a repository of content requires a substantial amount of effort. The question of the sustainability of the contributor community in Wikipedia has received attention from researchers~\cite{halfaker2013abs}, finding that there are technical and social barriers to entry.

The diversity of the contributor community has also been studied, as it is only possible to create a repository of the world's knowledge with the entire world taking part in creating that repository. In Wikipedia's case, the topic first received attention in 2011 after a survey of the community by the Wikimedia Foundation found that less than 15\% of the community were women~\cite{cohen2011gender}. Two additional studies further substantiated that this proportion was accurate~\cite{lam2011gender,makoshaw2013gender}. Studies have also shown how the lack of diversity affects peer-produced content such as lower amounts of content for movies in Wikipedia~\cite{lam2011gender}, how women are portrayed in Wikipedia biographies~\cite{wagner2021gender}, and that descriptions for services such as daycare and kindergarten could not be accurately described in OpenStreetMap~\cite{stephens2013gender}. When billions of people use the content that these communities produce every month, it is clear that solving these issues is essential, and attracting and retaining a more diverse community is one part of that.

It is unclear what solutions work when it comes to attracting and retaining newcomers at scale in peer production communities. Studies of Wikipedia have found that creating a welcoming environment where newcomers can have their questions answered in a friendly way has a positive impact~\cite{morgan2013tea,morgan2018teahouse}, that awarding contributors leads to higher productivity and retention~\cite{gallus2017fostering}, and that providing newcomers with a structured socialization process see them more likely to stick around and contribute~\cite{li2020wikied}. On the other hand, a gamified approach to teaching newcomers how to contribute received positive feedback from participants but did not affect the number of contributions~\cite{narayan2017}. These proposed solutions have only been available in a single community (English Wikipedia), and only two have been evaluated in controlled experiments.~\cite{morgan2018teahouse,narayan2017}.

In this paper, we propose the Newcomer Homepage as a possible solution for attracting and retaining newcomers in peer production communities. It is a central place where they can get a conceptual understanding of how peer production works, learn that they can contribute, and find help when they have questions. A central part of the homepage is Newcomer Tasks. This module helps solve the problem of finding opportunities to contribute by providing access to Wikipedia articles that require improvement and that are within specific topics of interest. When a newcomer selects an article to work on, they also receive guidance on completing their contribution.

In summary, this paper's contributions are:

\begin{itemize}
    \item We lay out the design of the Newcomer Homepage and show how it was built in accordance with research on the need of newcomers in Wikipedia~\cite{WMFNEE2017} and the Legitimate Peripheral Participation framework by Lave and Wenger~\cite{lw1991cop}.
    \item We explain how we developed, tested, and refined these solutions in collaboration with small and medium-sized Wikipedia communities.
    \item We evaluate the effect of the Newcomer Homepage and Newcomer Tasks through a large-scale controlled experiment spanning 27 Wikipedia communities, finding small overall changes in participation patterns but variations in effects for certain subgroups, and show how this has resulted in both existing and future changes to the homepage experience.
\end{itemize}

We will discuss the implications of our design process and experiment results for both Wikipedia in particular and peer production communities in general. Next, we situate this work more closely in relation to existing research.

\section{Related Work}
\label{sec:related}

\subsection{Peer Production Communities}
\textit{Peer production} is a term coined by Benkler~\cite{benkler2002coase} that refers to a system where many contributors make small contributions that come together to form a larger whole. Famous examples of this phenomenon are free/libre open source software (FLOSS), OpenStreetMap~\cite{haklay2008osm} (OSM), and the encyclopedia that is the focus of this paper.

There is an active contributor community behind all three examples of peer production we have listed, and the challenges of sustaining that community have been widely studied in the literature. The community of the English Wikipedia, the most prominent and well-studied edition of Wikipedia, grew exponentially from 2004 to 2007 before abruptly beginning to decline. Halfaker et al. studied this ``rise and decline'' pattern in their 2013 paper~\cite{halfaker2013abs}, finding that it correlated with an increase in rejection of what appeared to be good faith contributions from newcomers, leading to those newcomers leaving. Their study also found that the number of community norms and policies rose in tandem with the rising rejection of newcomers.

Most studies of Wikipedia have focused on English Wikipedia~\cite{okoli2012wikipedia,okoli2014wikipedia}, which means that the findings might not generalize to other communities. TeBlunthuis et al.~\cite{teblunthuis2018rise} replicated the analysis of Halfaker et al. using a sample of 740 wikis from Wikia.\footnote{Since renamed to Fandom: \url{https://about.fandom.com/about}} Their study found similar patterns and explanations for them, suggesting that this is a general trend in peer production communities and a challenge for their sustainability.

We add to this part of the literature in two ways: 1) we study a system that aims to help sustain peer production communities by attracting and retaining newcomers, and 2) we do not study English Wikipedia. Instead, we report findings from a controlled experiment with data from 27 non-English Wikipedia wikis.

\subsection{Socializing Newcomers}

A second line of research looks at the process of socializing newcomers, or in other words, how to introduce them to community norms and processes. One approach to socialization has been to design a friendly environment where newcomers can get answers to questions, and this exists in the English Wikipedia in the form of the Teahouse~\cite{morgan2013tea}. The Teahouse design involved a set of social norms for answering questions, and as the community followed these, it introduced a reinforcing circle of question-answering behavior~\cite{morgan2018welcome}. A controlled experiment with invites to the Teahouse showed that these resulted in increased retention of newcomers~\cite{morgan2018teahouse}.

Besides the formalization of how to answer questions, the Teahouse does not prescribe a formal process for socializing newcomers. The Wikipedia Education Program takes a different approach. Students in the program start by familiarizing themselves with Wikipedia norms before making simple and uncontroversial contributions, such as copy edits, and then move to more complicated tasks. Li et al.~\cite{li2020wikied} studied the effects of this process, finding that participants in the program made higher-quality contributions and were more likely to be retained as contributors relative to similar cohorts of other newcomers.

There are two other areas of research on socialization of newcomers in Wikipedia that are relevant to the current work. The first is mentoring newcomers, where Musicant et al.~\cite{musicant2011mentoring} reported mixed success in the program, highlighting that communication between a mentor and their mentee was missing critical aspects of good mentoring. The second area is introducing newcomers to the process of editing Wikipedia in a gamified and scaleable way, made available in Wikipedia through the Wikipedia Adventure. A study of this approach found that participants positively evaluated the tutorial, but it did not significantly impact subsequent newcomer contributions~\cite{narayan2017}.

Outside of Wikipedia, a study by Steinmacher et al. of barriers to newcomer contributions to open source software projects~\cite{steinmacher2018flosscoach} identified both finding answers to questions and finding a mentor as barriers. They developed a web portal that organizes information about projects and guides a newcomer through their first steps toward a successful contribution. They also suggest community guidelines, such as providing newcomers with mentors and ensuring that newcomer questions get answered quickly.

A study of newcomer contributors to OpenStreetMap~\cite{dittus2017private} found that verbal rewards and immediate feedback had a powerful effect on newcomer retention. In a study focused on ``mapping parties'', in-person events where people contribute to OSM, Hristova et al.~\cite{hristova2013osm} reported that these positively impacted less skilled contributors. However, they also report very low retention except for highly experienced contributors. Analysis of newcomer retention for teams who work on OpenStreetMap in areas where humanitarian efforts are needed~\cite{dittus2016analysing,mahmud2022revisiting} found that increased participation does not necessarily result in increased retention. Instead, teams with appropriate structures for long-term engagement and socialization were the ones who had better retention.

Our work follows these lines of research by providing Wikipedia newcomers access to a mentor and documentation that can help introduce them to Wikipedia's concepts, processes, and policies. In this paper, we do not study the mentoring process but focus on the system's general utility. We will return to the topic of mentorship when discussing our results and future work.

\subsection{Community Diversity}

Our work also relates to the diversity of the contributor community. As mentioned in the introduction, the issue of the lack of gender diversity in the Wikipedia community came up in 2011 after a community survey revealed that less than 15\% of the respondents were women~\cite{cohen2011gender}. This proportion has been further substantiated using another source of data from English Wikipedia~\cite{lam2011gender} and a study that combined survey responses with demographic data~\cite{makoshaw2013gender}.

The lack of gender diversity in the community is reflected in the content that it produces. Gender differences show up in the length of articles about movies in English Wikipedia~\cite{lam2011gender} and for information about places like kindergarten/daycare centers in OSM~\cite{stephens2013gender}. Regarding the gender diversity of biographies in English Wikipedia, research has found that Wikipedia has longer articles but is more likely to be missing biographies about women compared to the Encyclopædia Britannica~\cite{reagle2011gender}. The development of the gender diversity in biographies in Wikipedia can be monitored through the Wikidata Human Gender Indicators~\cite{klein2016whgi}, and a study has found that the coverage is improving over time~\cite{konieczny2018gender}. A study of specific projects to increase coverage in the English Wikipedia found them very successful~\cite{halfaker2017keilana}.

In order to attract and retain a diverse community, it is essential to create an environment that is welcoming and friendly. As discussed above, the English Wikipedia Teahouse aims to do that, but it is only a tiny corner of a vast community. One study pointed out how women who contribute to Wikipedia must carefully navigate the spaces they participate in to feel safe~\cite{menking2019people}. In another study examining why contributors to peer production might want to use privacy-preserving tools, interviewed Wikipedia editors reported threats of harassment as a key motivator~\cite{forte2017privacy}.

Research has also found a need for more geographic diversity in Wikipedia's content. Studies have found a geographic bias towards certain regions based on the wiki's language~\cite{HechtCT2009Bias}, for example in the form of a definitive lack of content about Africa in the English Wikipedia~\cite{graham2014geography}. Within a given country, e.g. the United States, geographic bias also shows up in the form of rural/urban bias~\cite{johnson2016}, with articles about rural areas being of lower quality relative to urban areas.

Studies of FLOSS communities identified barriers to entry, as mentioned earlier~\cite{steinmacher2018flosscoach}. In their study of barriers to entry in FLOSS communities~\cite{mendez2018open}, Mendez et al. found that these barriers interact with gender and disproportionately affect contributors who identify as women. Participation is not a question of motivation or ability but instead about these contributors facing significantly more barriers to entry. Ford et al. also found this interaction between barriers and gender in their study of participation on StackOverflow~\cite{ford2016paradise}. These findings indicate that communities cannot fix these problems through a small number of incremental improvements. Instead, they need more significant initiatives if they are serious about fixing things.

While the interventions we describe in this paper do not directly attempt to increase diversity, they aim to broadly encourage contribution and help newcomers avoid community resistance. These services are likely to be particularly helpful for underrepresented groups.

\subsection{Contribution Opportunities}

While large bodies of work created by peer production generally contain many opportunities to contribute, it is often unclear to newcomers how to begin doing so. This line of research is also related to socializing newcomers, as discussed above. We will cover two topics in this area: 1) understanding how peer production works, and 2) specific opportunities to contribute.

Understanding how one goes about contributing to Wikipedia is not apparent. Reboot and the Wikimedia Foundation's study of newcomers on the Czech and Korean Wikipedia~\cite{WMFNEE2017} reported that understanding how Wikipedia worked was a big challenge. A study by Hargittai and Shaw~\cite{hargittai2015gender} found that Internet skills and gender strongly link to whether someone will contribute, with highly-skilled men being the most likely.

Once a newcomer is on the site, finding opportunities to contribute might also be challenging. Recommender systems can alleviate this, as Cosley et al. showed in their study of SuggestBot~\cite{cosley2007sbot}. They found that users were likelier to edit articles related to those they had already edited than articles chosen randomly.

These issues are also barriers to contributions in other peer production communities. One of the guidelines for FLOSS communities suggested by Steinmacher et al.~\cite{steinmacher2018flosscoach} is to have a newcomer portal that explains how to contribute and points newcomers to easy tasks they can start with. In their previously mentioned StackOverflow study~\cite{ford2016paradise}, Ford et al. found a lack of easy questions to answer a barrier to newcomer participation. Another study of FLOSS communities that focused on one-time contributors~\cite{lee2017understanding} reported that they did not find additional opportunities to contribute. Secondly, they did not understand the processes that govern contributions in the project they were trying to contribute to.

In this paper, we describe a system that provides newcomers with a central place to learn how Wikipedia works to tackle the abovementioned conceptual problems. We also provide them with opportunities to contribute through Newcomer Tasks, which shows them articles needing improvement, similar to SuggestBot, but where the tasks are specifically newcomer-friendly.

\subsection{Design Motivation and Framework}

In their influential book ``Situated Learning'', Lave and Wenger describe the framework of Legitimate Peripheral Participation~\cite{lw1991cop}. We use this framework as guidance in our design, specifically focusing on three elements of their framework that we have also described above: access to participation, access to mentors, and access to learning materials. In the next section, we will make two connections between this framework and the results from the New Editor Experiences research mentioned above~\cite{WMFNEE2017}. First, the need for ``progressive pathways to editing'', which corresponds to Lave and Wenger's ``access to participation''. Second, the research reported on the challenge of understanding how Wikipedia works, and we will connect that to ``access to mentors'' and ``access to learning materials''. This brings us to the next section, where we will go into more detail about how we designed the Newcomer Homepage and how we came to make critical decisions in that process.

\section{Design}
\label{sec:design}

This section describes key design decisions behind the Newcomer Homepage and Newcomer Tasks. We start by focusing on how the design enables a newcomer to understand what participation means and provides access to the central activity in the community of practice. Secondly, we cover how the design needs to have a flexible user interface and allow for incremental improvements. Third, we show how the system addresses two essential needs from Lave and Wenger's Legitimate Peripheral Participation framework: access to mentors and sponsorship, and access to artifacts and learning materials~\cite{lw1991cop}. We wrap this section by describing the collaborative design process focusing on a small number of partner communities and how we designed for experimentation and iteration from the start.

\subsection{Access to and Understanding of Participation}

\begin{figure}
    \centering
    \includegraphics[width=0.9\textwidth]{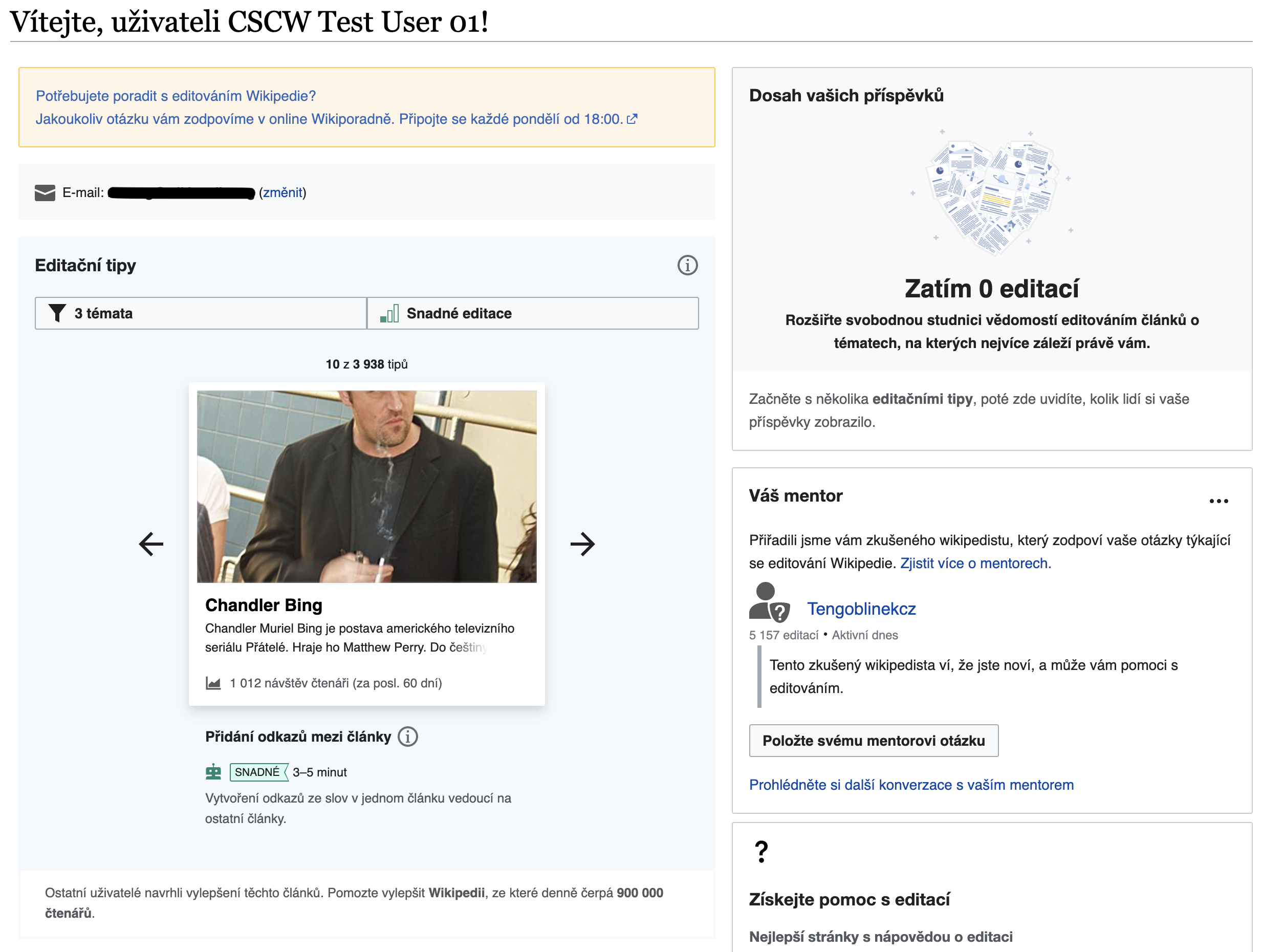}
    \caption{The desktop version of the homepage in Czech Wikipedia for the user ``CSCW Test User 01'', as reflected in the welcome greeting by username at the top. The Newcomer Tasks module (``Editační tipy'', English: editing tips) is the main user interface element on the left, and the Impact, Mentor, and partly hidden Help modules top to bottom on the right.}
    \label{fig:desktop_homepage}
\end{figure}

The Newcomer Homepage, shown in Figure~\ref{fig:desktop_homepage} (in Czech, the interface is localized to the local Wikipedia language), gives newcomers a central place to learn how Wikipedia works and that they can participate by editing. This concept is motivated from two key findings from the New Editor Experiences research conducted by Reboot and the Wikimedia Foundation in 2017~\cite{WMFNEE2017}. In that research, newcomers reported not always understanding that almost anyone can edit the encyclopedia (finding number 5). Secondly, they struggle with understanding the policies and participating in a way that does not violate them (finding number 8). We will discuss the latter in more detail in Section~\ref{subsec:mentor_learning} and instead turn our attention to how the Newcomer Homepage focuses on participation.

The Newcomer Homepage also provides newcomers with easier access to participate in what is arguably the core activity of Wikipedia: editing articles. In their work on understanding factors that lead to promotions to administrators in the English Wikipedia community~\cite{burke2008mop}, Burke and Kraut found that contributing to articles was consistently referred to in the promotion discussions. The New Editor Experiences research also found that newcomers need ``progressive pathways to editing'' (findings 7 and 9).

The part of the Newcomer Homepage that provides access to article editing is called ``Newcomer Tasks''. It is a central place for newcomers to find opportunities to edit and a place they can return to easily. The community has categorized the articles shown in this module as needing improvement, a similar approach to that of SuggestBot~\cite{cosley2007sbot} in order not to suggest mostly complete articles. Progression from easier to more challenging edits is made possible by changing the type of tasks shown, with some tasks categorized as ``Easy'' (e.g. copy editing, adding links), ``Medium'' (e.g. adding references), and ``Hard'' (e.g. expanding short articles).

Newcomers can tailor the list of articles by choosing their topics of interest. In their SuggestBot study, Cosley et al. found that randomly selected articles were much less likely to be edited compared to articles similar to those a user had shown interest in by editing. The Newcomer Homepage provides users with tens of topics enabled by the topic models in the ORES scoring service~\cite{halfaker2020ores}.

\subsection{User Interface Flexibility}

The Newcomer Homepage must support modern platforms, meaning it must function just as well on a mobile device as on a desktop computer. Most of Wikipedia's readership has come through mobile devices for many years~\cite{brown2015wpmobile}, while the site continues its struggle with being a ``child of the desktop internet''~\cite{economist2021wp20}. Since most account registrations on some wikis are on mobile (e.g. Arabic and Bangla), the homepage was designed from the start for mobile and desktop web. While there are Wikipedia apps for Android and iOS, the homepage is not available on these as they have focused on the reader experience, meaning limited editing features have been available. At the time of writing, there are no specific plans to bring the homepage to these apps.

Secondly, the Newcomer Homepage is module-based. This allows for extending it with new modules if new features are needed, removing old modules that are no longer useful, modifying modules to improve their functionality, and rearranging the layout of the modules on the page to emphasize specific actions. While not a focus of this study, the homepage has undergone several such changes since its deployment in May 2019.

\subsection{Providing Access to Mentors and Learning Materials}
\label{subsec:mentor_learning}

Two additional parts of Lave and Wenger's Legitimate Peripheral Participation framework appear on the Newcomer Homepage in specific modules: access to a mentor and learning materials. Newcomers can access a mentor through the Mentor Module, the second module from the top on the right-hand side in Figure~\ref{fig:desktop_homepage}. Experienced editors sign up to become mentors on a specific page on the wiki, where they also provide a welcoming message that will be shown in the module. At account registration, a newcomer is assigned to an active mentor. If the newcomer needs help they can use the module to ask their mentor a question, which is then posted to the mentor's user talk page.\footnote{A user talk page is a page associated with a specific user that is used for communicating with that user.} Mentors have guidance documentation available to help them respond to newcomer questions in a helpful way, fashioned after a similar approach on the English Wikipedia's Teahouse~\cite{morgan2013tea}. There is also functionality available to mentors to adjust how many mentees they would like to be assigned, to set themselves as unavailable (meaning they will not be assigned new mentees), or to take over a specific user as their mentee (e.g. when someone hosts a live event and wants to have all participants as their mentees).

The Help Module on the Newcomer Homepage provides access to pages where a newcomer can learn more about the norms and policies of the wiki, which we previously mentioned the New Editor Experiences research found to be a key challenge for newcomers, and as discussed in the related work is a frequent barrier to entry in peer-production communities. The Help Module shows five links to ``How to'' pages, chosen in consultation with the community. At the time of writing, the five ``How to'' links are about writing a good article, editing a page, adding an image, editing a citation, and creating a new article.

\subsection{Designing in Collaboration with Communities}
\label{subsec:design_collab}

The Newcomer Homepage has been designed and evaluated in collaboration with the wiki communities. We employed two strategies for this that are worth discussing in more detail: 1) Working closely with a small set of partner communities, and 2) Employing an experienced member of each community to act as a liaison.

The Newcomer Homepage was designed with a focus on small- to medium-sized wikis where there is an opportunity to grow the community, with the knowledge that the more established and larger communities could adopt the features later. This strategy differs from what has been a common thread in the research literature, where studies have focused on the English Wikipedia and then recommended that findings from that community be propagated to other communities. The English Wikipedia is an outlier on many dimensions, meaning findings from that community might not transfer well elsewhere. Secondly, we were planning to run live experiments with our features, and some of the larger communities, like the English Wikipedia, have developed resistance towards that.\footnote{See ``Wikipedia is not a laboratory'': \url{https://en.wikipedia.org/wiki/Wikipedia:NOTLAB}} As a result, we sought to create working relationships with communities open to collaboration and running experiments.

Two partner wikis, Czech and Korean, were inherited from the New Editor Experiences research. We invited two more, Vietnamese and Arabic, aiming to expand the cultural and geographical diversity. We shared design sketches with these partner communities, and feedback from them was gathered and translated by the community liaisons employed part-time by the Wikimedia Foundation. This setup enables a central point of contact known by the community, ensuring that conversations occur in the local language, and it aims not to exhaust volunteer resources.

When a feature was ready for testing, it would be deployed and tested locally by the partner communities first, particularly the community liaison. This has allowed for discovery and fixes to issues specific to that community, which are often related to localization (e.g. short phrases in English might be longer in other languages and therefore overflow the allotted space).

The central point of contact also enables the community to provide feedback after deployment in case new issues arise. For example, users who signed up to be mentors might take a break or leave the community, or there might be empty categories of tasks. Some of these issues get fixed through software patches by the development team (e.g. user interface issues). In contrast, others that have been recurring have resulted in the development of maintenance tools available to experienced community members to enable them to configure and maintain the system on their own.\footnote{\url{https://www.mediawiki.org/wiki/Special:MyLanguage/Growth/Community_configuration}}

\subsection{Planning for Experimentation}
\label{subsec:plan_for_exp}

The Newcomer Homepage has extensive instrumentation and data gathering to allow us to understand how users interact with it. Since the feature's success was uncertain, we wrote a measurement plan with several questions about user behavior (e.g. what proportion of newcomers visit their homepage within 48 hours of registration?) Those questions were then translated into instrumentation requirements so that data could answer them. Said instrumentation data was retained in accordance with the Wikimedia Foundation's data retention guidelines~\cite{WMFdataretention}.

\section{Evaluation}
\label{sec:evaluation}

In this section, we report our findings from a large scale controlled experiment with the Newcomer Homepage and the Newcomer Tasks module. This experiment ran from 1 February to 1 May 2021. Figure~\ref{fig:desktop_homepage} shows the module; it is the central one with the heading ``Editační tipy'' (Eng: editing tips).

Since deployment, we have run many controlled experiments with the Newcomer Homepage and the Newcomer Tasks module. In this paper, we focus  on the findings from this specific experiment for several reasons:

\begin{enumerate}
    \item It was a period where both the homepage and the module were feature-stable.
    \item They were deployed to many wikis before and during this period, enabling a more precise estimate of the effect of the features.
    \item The wider deployment means the results should also have broader applicability.
\end{enumerate}

This paper does not evaluate the Mentor module and its associated mentorship for several reasons. First, the feature has undergone much development to make it easier to maintain a community of mentors and have them interact with their mentees. Secondly, we have not run any experiments specific to this feature. Moreover, and most importantly, mentorship and socialization in peer production communities is a topic that warrants research by itself, meaning we plan to return to this in a future study.

\subsection{Datasets, Metrics, and Methods}
\label{subsec:datasets_metrics_methods}

Our dataset used for this analysis contains 244,060 accounts registered on one of 27 wikis during the experiment. A complete overview of the Wikipedia editions in the dataset and the number of accounts from each can be found in Appendix~\ref{app:dataset}.

During the experiment, an account was randomly assigned to the Treatment group (with 80\% probability) or the Control group at the time of registration. While a 50/50 split between these groups would provide us with higher statistical power, we chose this 80/20 split because our experiments up to that point had indicated a positive impact of the features. It would therefore be beneficial to give the features to a greater number of users. The increased size of the treatment group has also enabled us to run controlled experiments within that group to test variations of the Newcomer Tasks module. However, we do not report on these in this paper.

We excluded several types of accounts from the dataset. ``Autocreated accounts'', which are accounts created by the system for visiting users already registered on another Wikipedia project, were excluded. The same goes for accounts created by others, such as an administrator or event coordinator. Accounts known not to be eligible for the experiment, such as those registered by development team members, were also excluded. 

We also excluded registrations made through Wikipedia's API. These are generally either iOS or Android mobile app registrations, and the Newcomer Homepage is currently unavailable in the Wikipedia mobile apps. Lastly, although only new users get the features automatically, all users can manually turn them on or off in their account preferences. We excluded any users who did so, as this breaks the expectations of the random assignment.

We used data from two additional sources to extend our dataset of registrations. The first is event data captured at the time of registration, which tells us whether the registration occurred on Wikipedia's desktop or mobile web version. It also lets us determine if the user was editing before registration because the URL contains parameters to restart the Wikipedia editor.

The second additional data source is the Welcome Survey, a feature developed to understand newcomer motivation and previous experience with Wikipedia, potentially to personalize their newcomer experience.\footnote{\url{https://www.mediawiki.org/wiki/Special:MyLanguage/Growth/Personalized_first_day/Welcome_survey}} It was first deployed to the Czech and Korean Wikipedia in November 2018 and is at the time of writing a single-page survey that asks three questions:

\begin{enumerate}
    \item Why did you create your account today?
    \item Have you ever edited Wikipedia?
    \item Are there other languages you read and write in?
\end{enumerate}

The survey is shown to users directly after account registration. It is not required to answer the survey to finish account registration; users can skip it by clicking a link. We collected from this survey whether a user skipped or completed it and what their answers to the questions were if they completed it.

The metrics that we use will focus on a newcomer's editing activity. We have a specific set of four metrics that we use and report on:

\begin{description}
    \item[Activation] A newcomer is \textit{activated} if they edit within 24 hours of registering their account. We chose this 24-hour window based on an analysis of accounts registered in 2016 on the Czech and Korean Wikipedia who edited within one year of registration. In that analysis, we found that 85\% of Czech accounts and 91\% of Korean accounts edit on the first day. If we were to extend this to one week, we would only add about 5\% of accounts, and thus we decided to keep the 24-hour window.
    \item[Retention] An activated newcomer is \textit{retained} if they return to edit again on another day within two weeks. This period was also chosen based on the data used for defining activation. With the Newcomer Homepage aiming to affect newcomers early in their life cycle, this two-week window seeks to balance that focus while capturing most newcomers who will return to edit. For example, in the 2016 data from Czech Wikipedia, extending this window from two to four weeks would increase retention from 22.2\% to 24.5\%. Increasing the window is associated with an additional delay in data availability. We decided that the marginal gains did not counterbalance the increased delay and kept it at two weeks.
    \item[Productivity] A newcomer's \textit{productivity} is measured by the number of edits they make during a specific timeframe. We will focus on edits made across the entire ``newcomer period'', meaning the 15 days spanning both the activation and retention periods.
    \item[Revert rate] A newcomer's \textit{revert rate} is the proportion of edits they have made that have been reverted, meaning they were undone, within 48 hours of the edit being made. This 48-hour window is chosen based on Geiger and Halfaker's 2013 paper~\cite{geiger2013levee} that found that almost all reverts on the English Wikipedia are done within 48 hours.
\end{description} 

Regarding newcomer retention, we will also use two additional time windows to compare with the results of Morgan and Halfaker's 2018 paper studying the retention of newcomers invited to the English Wikipedia's Teahouse~\cite{morgan2018teahouse}. They used three different windows for retention and two different edit thresholds, finding that the Teahouse invite had a significant impact on short-term retention (3--4 weeks after registering) when a single edit was needed to be defined as ``retained'', and a significant impact on long-term retention (2--6 months) when at least five edits were needed. Given the proximity to our existing two-week window, our finding described above that increasing the window to four weeks does not capture a significant additional number of retained newcomers, and that the selection criteria for receiving the Teahouse invite are significantly different from our definition of an ``activated newcomer'', we will skip the 3--4 week window and instead focus on the longer term windows used in their paper (1--2 months and 2--6 months). We will keep the one edit threshold to keep the two additional windows comparable to our standard metric.

We will focus on edits made to the Article and Article talk namespaces (often referred to as ``Main'' and ``Talk'', respectively) because the Newcomer Tasks module invites newcomers to edit articles. There are many other namespaces in Wikipedia where activity can take place. For example, the User and User talk namespaces, where users provide information about themselves (User) and discuss with each other (User talk), or the Wikipedia and Wikipedia talk namespaces, where policies, guidelines, help information, and discussions about these take place. Finally, as one goal of the Newcomer Homepage and Newcomer Tasks is to enable newcomers to make positive contributions to the wiki, we will limit activation, retention, and productivity to what we define as \textit{constructive edits}, meaning edits that were not reverted within 48 hours of being made (see ``Revert rate'' above for the justification for this 48-hour threshold).

\begin{figure}
    \centering
    \includegraphics[scale=0.3]{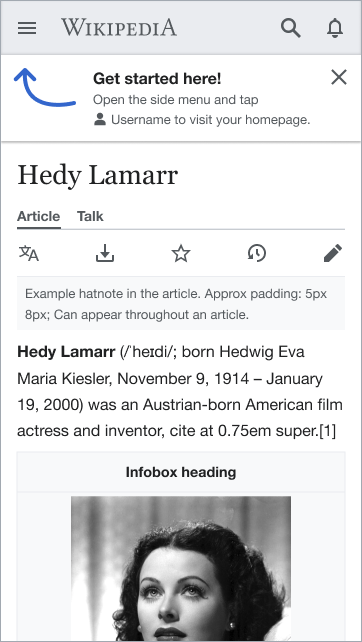}
    \caption{Mobile notification guiding a newcomer to the homepage when viewing an article after registration (shown in English, the message is localized to the Wikipedia language).}
    \label{fig:mobile_discovery}
\end{figure}

\begin{figure}
    \centering
    \includegraphics[width=0.8\textwidth]{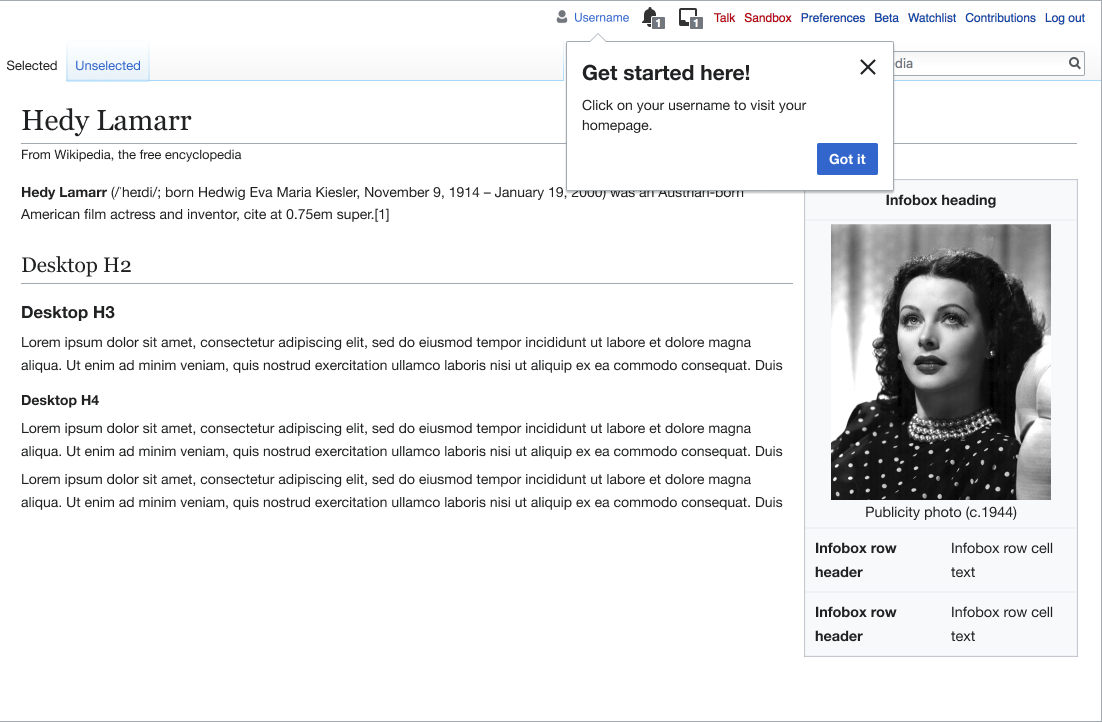}
    \caption{Desktop notification guiding a newcomer to the homepage when viewing an article after registration (shown in English, the message is localized to the Wikipedia language).}
    \label{fig:desktop_discovery}
\end{figure}

Our experiment is an example of ``encouragement design'' in that we encourage newcomers in the treatment group to visit the homepage and make suggested edits. Two examples of this encouragement are shown in Figures~\ref{fig:mobile_discovery} and \ref{fig:desktop_discovery}, and the messages are localized to the language of the Wikipedia where they are shown. In addition to these notifications, there are multiple links to the homepage. For example, the ``Username'' link in the set of links in the top right part of the desktop user interface (seen in Figure~\ref{fig:desktop_discovery}), in a notification shown after responding to the Welcome Survey, and from the list of user contributions if they have not made any contributions yet.

We analyze our experiment using two approaches. First is the ``Intent-to-Treat'' (ITT) approach, where we learn whether an invitation to the homepage results in significant differences. For this analysis, we use hierarchical regression models (also referred to as multilevel or mixed-effects models) with the wiki as the grouping variable to allow for wiki-related variation in baselines. The formula for this model is shown below in Equation~\ref{eq:itt_model} (adapted from Gelman and Hill~\cite{gelmanhill2006}), where $y_i$ is the outcome (e.g. activation or productivity) for a given user, $\alpha_{j[i]}$ is the intercept (baseline) that can vary for each wiki $j$, and $\beta_1$ and $\beta_2$ are the estimated coefficients for the treatment group assignment ($Group_i$) and platform of registration ($Platform_i$).

\begin{equation}
\label{eq:itt_model}
y_i = \alpha_{j[i]} + \beta_1 Group_i + \beta_2 Platform_i + \epsilon_i
\end{equation}

Activation and retention are binary outcomes, so we use a logistic regression model. Productivity is a count variable with a non-linear distribution, so we use a zero-inflated negative binomial regression model. Revert rate is a proportion where some users have \textit{some} of their edits reverted, some have \textit{all} reverted, and some have \textit{none} reverted. We use a zero-one-inflated beta regression model to model the revert rate, as it considers all three of those possibilities. Our ITT analysis was done in R, using the lme4~\cite{lme42015} package for our activation and retention models, and the brms~\cite{brms2017,brms2018} package for our productivity and revert rate models.

When reporting results from the ITT analysis we will focus on population-level effects, meaning that we estimate a coefficient across the whole dataset. Hierarchical regression models can also have group-level effects where a coefficient (e.g. $\beta_2$) is allowed to vary by group; it becomes a combination of population-level and group-specific estimates. In our case, this would mean that we can have wiki-specific coefficients.

We have also fitted models with group-level effects in our analysis and found that these tend to be challenging to discover or do not significantly improve fitness. In our experience, they require several relatively large wikis in the dataset and substantial variation between the population-level and wiki-specific estimates. Generally, the only place where we consistently find significant wiki-specific effects is the platform of registration (desktop or mobile). We have also tested our models for interaction effects and found no significant ones. In future work, we plan to look into how the characteristics of the wikis and variables affect our ability to discover wiki-specific effects.

Our second analytical approach estimates the effect of making suggested edits conditional on being invited to the homepage. Not all newcomers visit the Newcomer Homepage, and of those who do, only some proportion will go on to complete suggested edits. This self-selection and effort needed means we cannot compare these users against the entire control group, as many in the control group would also not complete suggested edits.

Estimating this conditional effect is done using two-stage least squares regression (2SLS)~\cite{angrist2009harmless}, and as our experiment design is similar to that of the Wikipedia Adventure~\cite{narayan2017}, we adopt their approach. In the first stage shown in Equation~\ref{eq:tsls_stage1}, we estimate the likelihood of making a suggested edit if invited to the homepage. These edits have a particular ``edit tag'' (labeled ``newcomer task'' in the system) making them easy to identify.

\begin{align}
    \label{eq:tsls_stage1} edited &\sim invited + Platform * wiki + \epsilon \\
    \label{eq:tsls_stage2} y &\sim \widehat{edited} + Platform * wiki + \epsilon
\end{align}

We control for the platform of registration and the wiki with a possible interaction effect between the two, thus allowing for per-wiki differences between desktop and mobile users and making the 2SLS models reflect our wiki-specific effects described earlier. In the second stage shown in Equation~\ref{eq:tsls_stage2}, we use the fitted values from the first stage as a predictor combined with the same control variables. The result is an unbiased estimate of the causal effect of making suggested edits conditional on being invited.

Making a suggested edit means editing a Wikipedia article, making these edits part of our previously described measurements. In the 2SLS analysis, we exclude these edits from the outcome measurements to not introduce bias in the analysis. Our 2SLS analysis was also done in R with the ivreg~\cite{ivreg2021} package.

\subsection{Ethical Considerations}

Our evaluation of the Newcomer Homepage and Newcomer Tasks involved running a large-scale controlled experiment. Doing so is not without ethical concerns, as we deliberately change the newcomer experience on the platform. While this work did not go through a formal review process, we did our best to ensure it was done well.

Every community had to decide to be part of the experiment before the features were deployed to that community. This decision was reached through the community's consensus process, typically involving a proposal posted in a community forum followed by a discussion amongst active members. Being part of the experiment also meant the community needed to translate the user interface and ensure that help pages were available. Through this process, the community also established points of contact with the experiment team in case issues arose.

We judged the risk to experiment participants to be low. The treatment group got access to a set of additional features, and there were no changes to the control group's experience during the experiment. Users in the treatment group were encouraged to visit the homepage and use its features. However, they were not required to do so, and using the features did not involve additional system privileges or benefits. While we did not inform users that they were part of the experiment, they could freely opt in or out by changing their user preference settings.

This experiment, and the others we have run with the Newcomer Homepage and Newcomer Tasks, were all time-limited. They have had publicly available experiment plans, which defined leading indicators we would monitor and evaluate during the first two weeks of an experiment. If these indicated that something was not working as it should, the experiment team would work to mitigate the issue.

\subsection{Overall Findings}
\label{subsec:findings}

We start by reporting our findings for the four outcome metrics: activation, retention, productivity, and revert rate. For the first three, we show results from both our Intent-to-Treat analysis and the two-stage least squares analysis.

\subsubsection{Activation}
\label{subsubsec:activation}

Our intent-to-treat analysis finds a \textit{small but significant increase} in constructive activation in the Article and Article talk namespaces. The estimates from the hierarchical logistic regression model are shown in Table~\ref{tab:itt_activation}, where we can see that the treatment group estimate is $0.036$, which is roughly equivalent to a 1\% increase in the odds of activation.\footnote{Per the ``divide by 4 rule'' from Gelman and Hill~\cite{gelmanhill2006}.} 

Our two-stage least squares analysis of activation finds instead that those who go on to make suggested edits are \textit{significantly less likely} to activate. Table~\ref{tab:tsls_activation} shows the estimates from that model, where we can see that having edited has a strong negative effect on the likelihood of activating. 

\begin{table}
    \centering
    \caption{User Activation -- ITT Model Estimates}
    \begin{tabular}{ l r r r }
        \toprule
        & \textbf{Estimate} & \textbf{Std. error} & \textbf{P-value} \\
        \midrule
        Intercept & -1.052 & 0.078 & $P \ll 0.001$ \\
        $Platform = Mobile$ & -0.388 & 0.041 & $P \ll 0.001$ \\
        $Group = Treatment$ & 0.036 & 0.012 & $P < 0.005$ \\
        \bottomrule
        \multicolumn{4}{l}{ $Cond. r^2=0.061$,
        deviance $D = 257 645.6$, null deviance $D_{null} = 259 842.8$}
    \end{tabular}
    \label{tab:itt_activation}
\end{table}

\begin{table}
    \centering
    \caption{User Activation -- 2SLS Model Estimates}
    \begin{tabular}{ l r r r }
        \toprule
        & \textbf{Estimate} & \textbf{Std. error} & \textbf{P-value} \\
        \midrule
        Intercept & 0.156 & 0.003 & $P \ll 0.001$ \\
        $Platform = Mobile$ & -0.070 & 0.002 & $P \ll 0.001$ \\
        $Edited = Yes$ & -0.343 & 0.109 & $P \ll 0.001$ \\
        \bottomrule
        \multicolumn{4}{l}{ $Adj. r^2=0.019$}
    \end{tabular}
    \label{tab:tsls_activation}
\end{table}

Combining these two findings, we see that the Homepage invitation and Newcomer Tasks do not substantially increase the probability that a newcomer makes their first edit. Part of the reason is that for newcomers who follow the invitation and make suggested edits, these edits replace those they would otherwise make, as shown by the negative estimate in the 2SLS analysis. Whether a peer production community benefits from newcomers changing their trajectory from doing non-prompted edits to other types of edits depends on many factors. We will dig into some of those factors in Section~\ref{subsec:morecontext}, exploring the effects of the newcomer context.

The estimate of the effect of the platform of registration (desktop or mobile) is the specific area where our models find significant per-wiki variation (we do not list the individual estimates for brevity). This phenomenon occurs consistently across our analyses of our controlled experiments. We know that the proportion of users who register on the mobile platform varies significantly from one Wikipedia to another. As a result, it is not surprising that the difference in probability of activation between platforms also varies. One might also hypothesize that there would be per-wiki differences in the effect of being invited. However, our work on these regression models has not found it significant enough to warrant inclusion in the ITT models, as determined by comparing the Bayesian Information Criterion (BIC) between models. The BIC of the ITT model reported in Table~\ref{tab:itt_activation} is 257,936, whereas if we add per-wiki variation, it becomes 257,970. This substantial difference in BIC makes the first model the preferable choice. The challenges of identifying and understanding wiki-specific effects is also a topic we will return to in the discussion section.

\subsubsection{Retention}
\label{subsubsec:retention}

The ITT model estimates for editor retention are shown in Table~\ref{tab:itt_retention}, where we can see that the effect of being invited to the Homepage ($Group = Treatment$) is not a significant predictor. These estimates indicate that the Newcomer Homepage and Newcomer Tasks do not fundamentally change why newcomers stick around.

Our 2SLS model also finds a similar result. Table~\ref{tab:tsls_retention} shows the model estimates, and having made suggested edits does not alter the likelihood of being retained. This means that the reduction in activation that we saw in the previous section appears to be a short-term phenomenon. In their analysis of the Wikipedia Adventure~\cite{narayan2017}, Narayan et al. reported a similar pattern of long-term null effects and a likely short-term reduction.

\begin{table}
    \centering
    \caption{User Retention -- ITT Model Estimates}
    \begin{tabular}{ l r r r }
        \toprule
        & \textbf{Estimate} & \textbf{Std. error} & \textbf{P-value} \\
        \midrule
        Intercept & -28.675 & 0.738 & $P \ll 0.001$ \\
        $Platform = Mobile$ & -0.256 & 0.028 & $P \ll 0.001$ \\
        $Group = Treatment$ & 0.032 & 0.031 & $P = 0.31$ \\
        $Activation = True$ & 25.603 & 0.738 & $P \ll 0.001$ \\
        $log(edits)$ & 1.183 & 0.191 & $P \ll 0.001$ \\
        \bottomrule
        \multicolumn{4}{l}{$Cond. r^2 =0.975$, deviance $D = 43571.6$, null deviance $D_{null} = 71827.6$}
    \end{tabular}
    \label{tab:itt_retention}
\end{table}

\begin{table}
    \centering
    \caption{User Retention -- 2SLS Model Estimates}
    \begin{tabular}{ l r r r }
        \toprule
        & \textbf{Estimate} & \textbf{Std. error} & \textbf{P-value} \\
        \midrule
        Intercept & -0.002 & 0.001 & $P = 0.18$ \\
        $Platform = Mobile$ & -0.003 & 0.002 & $P < 0.1$ \\
        $log(edits)$ & 0.148 & 0.001 & $P \ll 0.001$ \\
        $Edited = Yes$ & 0.024 & 0.042 & $P = 0.57$ \\
        \bottomrule
        \multicolumn{4}{l}{ $Adj. r^2=0.188$}
    \end{tabular}
    \label{tab:tsls_retention}
\end{table}

Extending the retention window to match the Teahouse retention paper as we described in Section~\ref{subsec:datasets_metrics_methods} also results in no difference between the Control and Treatment groups in an ITT analysis. The estimates from the regression models are left out for brevity as they are largely similar to those shown in Table~\ref{tab:itt_retention}. We also find that the registration platform stops showing a significant difference, leading us to leave it out of those models. This finding suggests that the negative correlation between platform and retention in Table~\ref{tab:itt_retention} is also a short-term effect.

The focus of many studies of Wikipedia, including ours, has been on how to \textit{improve} retention rather than how to \textit{understand} retention. Future studies focusing on statistical models and factors influencing newcomer retention in peer production can help mitigate this. Our models show that editing activity is positively associated with retention, as expected. At the same time, we also learn that newcomers on the mobile platform are less likely to be retained in the short term. Different strategies that adapt to the newcomer's context might be needed to retain them successfully. We will return to this topic in Section~\ref{subsec:morecontext} and bring it up in the Discussion section as future work focuses on improving the retention of activated newcomers.

\subsubsection{Productivity}

We use a zero-inflated negative binomial model fitted using a Bayesian approach in our ITT productivity analysis, as discussed in Section~\ref{subsec:datasets_metrics_methods}. For this model, we report estimated means, errors, and 95\% credible intervals, shown in Table~\ref{tab:itt_productivity}. As we can see in that table, the estimated mean for newcomers in the treatment group ($Group = Treatment$) is small and close to zero, and the 95\% credible interval spans across zero. This means we do not find evidence that being invited to the Newcomer Homepage and having access to Newcomer Tasks increases newcomer productivity.

In the same table, we can also see that registering on the mobile platform ($Platform = Mobile$) is associated with lower overall productivity (95\% CI $[-0.47, -0.03]$) and a higher likelihood of not making any edits at all (zero-inflation $Platform = Mobile$ 95\% CI $[1.49,4.58]$ estimated using an inverse logit function).

\begin{table}
    \centering
    \caption{User Productivity -- ITT Model Estimates}
    \begin{tabular}{ l r r r r}
        \toprule
        & \textbf{Mean} & \textbf{Est. Error} & \multicolumn{2}{c}{\textbf{95\% Cred. Int.}} \\
        \midrule
        Intercept & 0.53 & 0.11 & 0.31 & 0.74 \\
        $Platform = Mobile$ & 0.11 & -0.26 & -0.47 & -0.03 \\
        $Group = Treatment$ & -0.02 & 0.01 & -0.05 & 0.01 \\
        Zero-inflation intercept & -4.29 & 0.95 & -6.47 & -2.63 \\
        Zero-inflation $Platform = Mobile$ & 2.79 & 0.76 & 1.49 & 4.58 \\
        \bottomrule
    \end{tabular}
    \label{tab:itt_productivity}
\end{table}

\begin{table}
    \centering
    \caption{User Productivity -- 2SLS Model Estimates}
    \begin{tabular}{ l r r r }
        \toprule
        & \textbf{Estimate} & \textbf{Std. error} & \textbf{P-value} \\
        \midrule
        Intercept & 0.156 & 0.005 & $P \ll 0.001$ \\
        $Platform = Mobile$ & -0.014 & 0.006 & $P < 0.05$ \\
        $Edited = Yes$ & -0.536 & 0.147 & $P < 0.001$ \\
        \bottomrule
        \multicolumn{4}{l}{ $Adj. r^2=0.003$}
    \end{tabular}
    \label{tab:tsls_productivity}
\end{table}

Our 2SLS model estimates for productivity are shown in Table~\ref{tab:tsls_productivity}. Here we again find a substantial negative effect of having made suggested edits if one were invited to the homepage, similarly as we saw for activation. This again means that suggested edits appear to replace other contributions rather than augment them.

Our findings related to productivity suggest that we did not fundamentally change editing, and similarly to activation, we redirected attention from non-prompted edits toward Newcomer Tasks. While the Help module on the homepage provides newcomers with information about editing, they still have many things they need to figure out themselves. Adding more structure to the editing process might help reduce the effort needed to contribute, an approach we will discuss later.

\subsubsection{Revert rate}

We find no difference between the treatment and control groups regarding the probability of their edits being undone in our ITT analysis of revert rates. The estimates from our zero-one-inflated beta model are left out for brevity as the model estimates four outcome variables. For each outcome variable, we have an intercept and three independent variables (group, platform, and the number of edits made). Instead of listing all 16 estimates, we focus on how the estimates of two control variables are consistent with what we would expect and draw some conclusions about what our finding of no difference means.

While our usage of a zero-one-inflated beta model on revert rates is novel, the estimates from our model make sense. Mobile registration is associated with a higher likelihood of being reverted, something we would also see if we compared revert rates between desktop and mobile edits. We also find that the number of edits made relates to revert rates meaningfully. An increase in the number of edits made is associated with a lower likelihood of having all edits reverted, which we interpret to be related to Wikipedia warning and banning users who consistently make non-constructive edits~\cite{Geiger2010}. It is also associated with a lower likelihood of having none of your edits reverted. Since reverts are a part of Wikipedia's quality assurance process~\cite{Halfaker2009}, it makes sense that an increase in one's edit count correlates with an increase in the likelihood that someone in the community will contest at least one of those edits. Lastly, our model also estimates the mean and shape of the beta distribution, and making more edits is associated with a lower mean revert rate.

We interpret our finding of no difference in the revert rate and our finding of no difference in productivity as further substantiation that we have not fundamentally changed the editing experience. As we also mentioned in our summary of the productivity analysis, a more structured intervention that takes away the guesswork of understanding how to contribute constructively to Wikipedia could make a more substantial impact.

\subsection{Understanding the Newcomer Context}
\label{subsec:morecontext}

The findings from our ITT and 2SLS analyses motivate further investigation into the relationship between the newcomer's context at registration and the effects of our intervention. As described in Section~\ref{subsec:datasets_metrics_methods}, we have a dataset on whether a newcomer attempted to make an edit when they registered, together with survey responses that tell us more about the newcomers' goals and experience. We will focus on how this connects to our ITT analysis of newcomer activation and retention, showing how context can reveal critical insights for specific subpopulations.

\subsubsection{Editing at Account Creation}
\label{subsubsec:editing}

\begin{figure}[htp]
    \centering
    \includegraphics[scale=0.4]{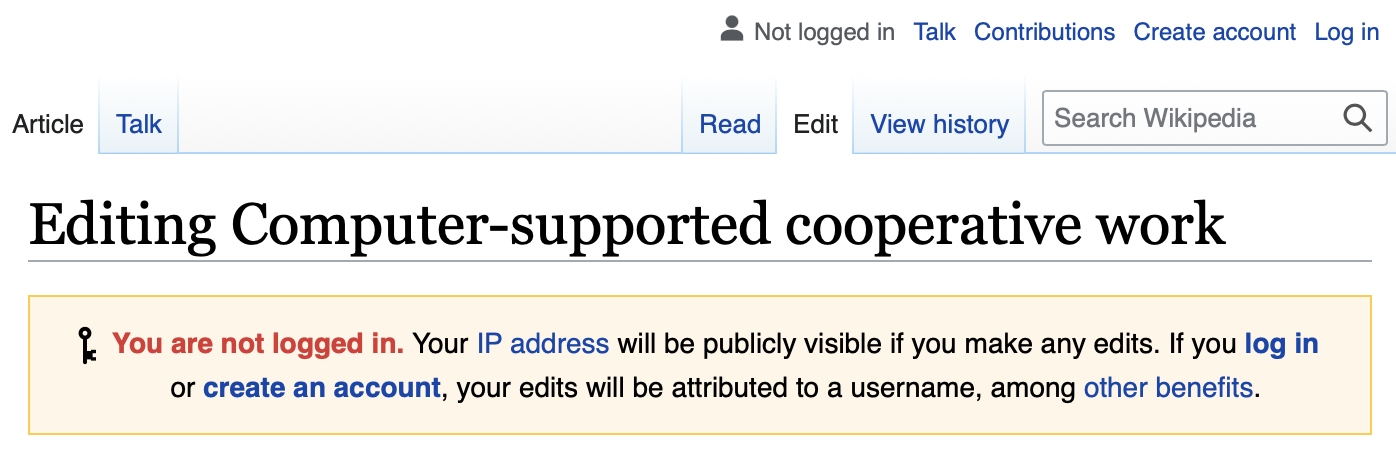}
    \caption{Warning message shown when trying to edit on the English Wikipedia without logging in }
    \label{fig:anoneditwarning}
\end{figure}

Users who try to edit a Wikipedia article without logging in will typically see a message warning them that they are not logged in, as seen in Figure~\ref{fig:anoneditwarning}. In 2014, the Growth team at the Wikimedia Foundation experimented with variations of this message and found conflicting results~\cite{WMFanonresearch}. A pre-edit call to action would increase registrations but decrease the number of completed edits. Unfortunately, the team was disbanded before this research was completed, as retold in Section~2.2 of Morgan and Halfaker~\cite{morgan2018teahouse}.

The features inviting newcomers to the homepage described in Section~\ref{subsec:datasets_metrics_methods}, such as the ones shown in Figures~\ref{fig:mobile_discovery} and \ref{fig:desktop_discovery}, might interrupt their workflow if they attempted to edit before registering their account. Upon successful registration, the user is taken back to the Wikipedia editor but might then be distracted from their editing task by our notification that they can click on their username to go to the homepage. In our dataset, 26.1\% of newcomers were editing at the time of registration, which means that if we negatively affect their experience, it can have a powerful impact.

\begin{table}[h]
    \centering
    \caption{User Activation and Editing -- ITT Model Estimates}
    \begin{tabular}{ l r r r }
        \toprule
        & Estimate & Std. error & P-value \\
        \midrule
        Intercept & -1.1781 & 0.0812 & $P \ll 0.001$ \\
        $Platform = Mobile$ & -0.7403 & 0.0424 & $P \ll 0.001$ \\
        $Group = Treatment$ & 0.0699 & 0.0149 & $P \ll 0.001$ \\
        $Editing = True$ & 0.9384 & 0.0239 & $P \ll 0.001$ \\
        $Group * Editing$ & -0.1136 & 0.0261 & $P \ll 0.001$ \\
        \bottomrule
        \multicolumn{4}{l}{$Cond. r^2 =0.104$, deviance $D = 252472.8$, null deviance $D_{null} = 259842.8$}
    \end{tabular}
    \label{tab:editcontext}
\end{table}

We find that users who were editing at registration saw a significant decrease in constructive article activation if they were also in the treatment group. The estimates from our ITT regression model that accounts for editing and the interaction between group assignment and editing are shown in Table~\ref{tab:editcontext}. We have also undertaken a 2SLS analysis, which found similar results and has been left out to save space.

As expected, editing at registration is strongly associated with a higher likelihood of activation. More importantly, we see that being in the treatment group is positively associated unless the user was also editing, where the decrease in the estimated interaction ($-0.114$) is almost double that of the estimate of being in the treatment group ($0.07$). We also found this effect in the 2SLS analysis, where editing at registration was associated with decreased activation, similar to what we reported in Section~\ref{subsubsec:activation}. In contrast, not editing was associated with no significant difference.

From this analysis, it seems clear that the Welcome Survey and the invitations to the Newcomer Homepage distract the user's attention, and they do not go on to complete the edit they had started. We have since then redesigned the user experience for users who are editing at registration such that they do not get information about the homepage until after they have completed their edit.\footnote{\url{https://phabricator.wikimedia.org/T310320}}

\subsubsection{Signing Up to Read Wikipedia}

The Welcome Survey described in Section~\ref{subsec:datasets_metrics_methods} provides several pre-filled options for answering two of the questions, and one of the responses for creating an account is ``To read Wikipedia.'' In an analysis a month after the first release of the survey,\footnote{\url{https://www.mediawiki.org/wiki/Special:MyLanguage/Growth/Analytics_updates/Welcome_survey_initial_report}} we found both high response rates (67\% in Czech and 62\% in Korean) and that a reasonably large proportion of responses answered the question this way (18\% in Czech and 29\% in Korean). We interpret these responses to suggest that the Newcomer Homepage and Newcomer Tasks can help newcomers learn that editing is possible and help them accomplish it.

The response rate in our dataset is 68.5\%, and 14.9\% of respondents said they signed up to read Wikipedia (10.2\% of all users). We removed respondents who were editing at registration but claimed they signed up to read as that is a contradiction, and these account for 2.1\% of all users. The remaining dataset consists of 19,847 respondents (8.1\% of all users).

The survey also asks whether users have edited before, with several different pre-filled responses available. In our work on this analysis, we found that the two responses that correspond to some variant of ``yes'' are associated with a substantially higher activation likelihood, and all the other responses fall together into a lower likelihood. Therefore, we treat the response to that question as a yes/no answer.

\begin{table}[h]
    \centering
    \caption{Activation of Users Signing up to Read -- ITT Model Estimates}
    \begin{tabular}{ l r r r }
        \toprule
        & Estimate & Std. error & P-value \\
        \midrule
        Intercept & -3.0996 & 0.1156 & $P \ll 0.001$ \\
        $Platform = Mobile$ &  -0.3114 & 0.0642 & $P \ll 0.001$ \\
        $Group = Treatment$ & 0.4296 & 0.0149 & $P \ll 0.001$ \\
        $Edited = True$ & 0.9017 & 0.0669 & $P \ll 0.001$ \\
        \bottomrule
        \multicolumn{4}{l}{$N = 19847$, $Cond. r^2  0.075$, deviance $D = 9075.7$, null deviance $D_{null} = 9289.6$}
    \end{tabular}
    \label{tab:readers}
\end{table}

We found a significant increase in activation for respondents who said they created their account to read Wikipedia. The estimates from our ITT model are shown in Table~\ref{tab:readers}. While the estimated effect is relatively large, it is worth mentioning that the baseline activation rate of these users is relatively low. In the control group, the activation rate of users who have not edited before is 3.8\%, compared to 9.9\% for those who have edited before. This activation rate is substantially lower than the overall activation rate in the control group, which is 23.1\%.

These results indicate that interventions like the Newcomer Homepage and Newcomer Tasks can help convert someone into a contributor in cases where they otherwise would not have made that transition. One of the challenges with affecting change in a community like Wikipedia is that participation requires self-selection and discovery. Someone new to the community might not know what possibilities exist. The homepage solves this problem by directing newcomers to a place where potential contributions are right in front of them.

We now turn our attention to discussing these findings in both a broader context of Wikipedia research as well as peer production communities in general.

\section{Discussion}
\label{sec:discussion}

We have described key design decisions behind the Newcomer Homepage and Newcomer Tasks and shown how these interventions can lead to positive and negative outcomes for newcomers in peer-production communities. Our metrics focused on newcomer participation in the form of making their first contribution (activation), returning to contribute again (retention), productivity (number of constructive contributions made), and an overall measure of the constructiveness of contributions (revert rate). In summary, our findings were:

\begin{description}
    \item[Activation] A small but significant increase in overall activation, and that the outcome depends on newcomer context. Our intervention appears to distract newcomers who were already in the process of contributing, but seems to support those who were not, and in particular those who did not create an account with an intention to contribute.
    \item[Retention] No difference in the retention rate; we find a strong correlation with the activity level on a newcomer's first day.
    \item[Productivity] No difference in the overall number of constructive contributions.
    \item[Revert rate] No difference in the proportion of contributions rejected by the community.
\end{description}

We have also described how we chose to continue with the parts of the intervention that appeared to work and change the parts that did not. In this section, we will consider the design, the findings, and our decisions in a broader context, as well as some of the limitations of our work and what opportunities for future work this study motivates.

\subsection{Scale of Intervention}

In their book ``Situated Learning''~\cite{lw1991cop}, Lave and Wenger describe how learners need particular types of access, with access to participation being top of the list. We also pointed to related work by Steinmacher et al.~\cite{steinmacher2018flosscoach} describing how access to documentation, mentorship, and easy newcomer tasks are barriers to entry in Open Source Software communities. We described how the Newcomer Homepage follows these guiding principles and how the Newcomer Tasks module enables access to editing articles, the central task in Wikipedia. Our experimental evaluation showed that we could guide some newcomers towards this task and increase the likelihood that they chose to participate, while we might be distracting others from completing their contribution.

We also found that the Newcomer Homepage and Newcomer Tasks did not fundamentally alter the obstacles that make editing Wikipedia difficult for newcomers. In order to do so, it might be necessary to make more dramatic changes to the editing process. The Wikimedia Foundation is currently experimenting with this in the form of a more structured editing experience for Newcomer Tasks where the newcomer makes decisions based on machine learning suggestions and gets specific prompts to write when writing is needed. This is a paradigm shift in the Wikipedia editing experience; newcomers are no longer let loose in the Wikipedia editor with the expectation that they translate concepts like ``add links to other articles'' into edits on their own. If this type of intervention is successful, it suggests that significant changes to how a contribution is made might be needed to make these interventions successful.

Do impactful interventions in peer production communities require substantial resources? We know of only one controlled experiment that has shown a significant impact on newcomer retention: the Teahouse on English Wikipedia~\cite{morgan2013tea,morgan2018teahouse}. The Teahouse project had access to significant resources and made design decisions that deviated from the community norms (e.g. that new topics would appear at the top of the page when the community standard was to put them at the bottom). The Newcomer Homepage is a multi-year project with dedicated management, design, engineering, QA, and research resources behind it. Peer production is often seen as a grassroots effort built from the ground up. It has successfully created encyclopedias, operating systems, and support tools like the automated vandalism tools used in Wikipedia. There might be specific problems that this incremental bottom-up approach is not suited to solve, with engaging and retaining newcomers possibly being one of those.

\subsection{Born \textit{and} Made}

Previous research has framed a question of whether ``Wikipedians'' (successful core contributors to Wikipedia) are \textit{born}, meaning their success is predetermined by skills and attributes acquired before they show up, or \textit{made}, meaning their success is a product of their experiences and learning after they sign up. In their 2005 paper ``Becoming Wikipedian''~\cite{bryant2005becoming}, Bryant et al. interviewed three central Wikipedia contributors who described going through a learning process after creating their account, which supports the ``made'' hypothesis. The 2009 paper ``Wikipedians are Born, not Made'' by Panciera et al.~\cite{panciera2009} took a quantitative approach to the question and showed that the highly prolific contributors were prolific from day one. Research that has shown an impact on newcomer retention, such as the already mentioned Teahouse work and Gallus's 2017 experiment~\cite{gallus2017fostering} with barnstars,\footnote{Barnstars are given to a user as a reward for good work in the English Wikipedia.} both used selection criteria that mean users were already ``made''.

Huang et al.~\cite{huang2015activists} add nuance to this picture, finding that volunteers in Change.org, a sizeable online e-petition platform, could be both born and made. Jackson et al.~\cite{jackson2018did} studied contribution patterns by non-registered and registered users in Zooniverse, an online citizen science platform. They found that several contributors who were active from the start had previously made contributions without logging in.

We also add nuance to this picture by using responses from the Welcome Survey, finding that some newcomers are born, but others can be made. Some ``newcomers'' show up highly motivated with previous experience and may even have registered while editing. These might go about their business regardless of our intervention, as they already know the system and what they want to be doing. Others have experience but might need help knowing where to contribute, which the Newcomer Homepage and Newcomer Tasks can help address. Finally, as the survey responses showed, some newcomers may sign up intending to read but end up being shifted onto the path of contributing by our interventions.

\subsection{Newcomer Contributions and Community Needs}

Our two-stage least squares analysis and investigation of activation for users who were editing at registration pointed to our intervention diverting contributor effort away from what they were doing. Whether this is a benefit or a drawback depends on many factors. A large and prosperous community might focus on maintaining quality, favoring semi-automated reviewing solutions that ease the burden but likely demotivate newcomers~\cite{halfaker2013abs,teblunthuis2018rise}. These communities might also seek to limit contributions; it is currently only possible to create new articles in the English Wikipedia with a registered account that is at least four days old and has made at least ten edits. Whether contributing should be possible without logging in is also a frequent topic of discussion, although research points to non-registered contributions often being valuable~\cite{anthony2009reputation} and that the net benefit for peer-production communities is positive~\cite{hill2021hidden}.

Smaller communities might be more open to all kinds of contributions as their goal might be to increase the amount of content, e.g. through translating high quality articles in the case of an online encyclopedia~\cite{urwikipedia2012}, or adding features in the case of open-source software. The Newcomer Tasks module is set up to allow any community to guide newcomers towards tasks that the community has labeled as valuable, whether adding new content or improving what is already there.

As mentioned above, some ``newcomers'' are highly motivated and experienced. If they are in the process of making a valuable contribution at the time of registration, we want our interventions to stay out of their way. Our analysis of users who were editing at registration in Section~\ref{subsubsec:editing} found a negative impact. We described how we redesigned the user experience so these users can return to contributing. However, suppose they are about to make a contribution likely to be discarded by the community. In that case, we could implement editing tools that assist them in improving the contribution, like the one developed by Chen et al.~\cite{chen2017community}

\subsection{Limitations and Future Work}

Our findings supporting that newcomers are both ``born'' and ``made'' have implications for the composition of peer production communities. Wikipedia's issue with gender bias in its community came to light in 2011~\cite{lam2011gender}. Studies have focused on understanding gender representation in the community~\cite{makoshaw2013gender}, how the content in the encyclopedia is affected, and how the community is hostile to women. Gender bias is not a problem unique to Wikipedia, with a study on OpenStreetMap showing examples of how resources valuable to women were dismissed~\cite{stephens2013gender}. Designing, implementing, and studying broader interventions such as ours that allow newcomers to discover all the possibilities to contribute, rather than continuing to select for newcomers who have already opted in to the current state of affairs, could play a crucial role in making a shift in who contributes and with that a shift in the breadth of content that is produced.

The Newcomer Homepage also provides access to a mentor and learning materials, two additional vital parts of Lave and Wenger's framework. Research on mentorship in Wikipedia~\cite{musicant2011mentoring} showed the mixed success of the informal mentorship systems in the English and German Wikipedia at the time. In more recent work studying the formal socialization process of the Wiki Education Program, Li et al.~\cite{li2020wikied} found that students participating in the program contributed more to the encyclopedia than similar cohorts of other newcomers. Students in the program were also more likely to continue contributing after program completion.

We did not study the mentorship or socialization aspects of the Newcomer Homepage closely in this paper. Instead, we are adding more structure to the overall newcomer experience by introducing elements of positive reinforcement, such as goals and badges. As part of analyzing the effects of these changes, we might also examine how access to a mentor affects newcomers.

Studying the effects of interventions across communities and subgroups is essential, so we can learn where successes and failures occur. We have reported on a large-scale experiment spanning a substantial number of communities in this paper. In our findings, we discussed how this can make discovering effects on smaller groups challenging. Some communities do not see a large influx of newcomers and therefore require a long time before it is possible to determine that there is a significant effect, provided there is one. As a result, studying larger communities is more attractive. However, those communities might have structures that dampen the effects of interventions, such as the increase in policies and guidelines described in the two papers studying the ``Rise and Decline'' of active contributors~\cite{halfaker2013abs,teblunthuis2018rise}. Neither the English nor the German Wikipedia participated in our experiments, but the features have rolled out to both wikis during 2021 and 2022. We will continue to work across larger and smaller communities and use statistical methods that pool datasets to discover effects in experiments.

There are opportunities to use additional datasets to enrich our understanding within a single peer production community and across multiple communities. Studies examining multiple peer production communities have often used them comparatively, such as how content or policies differ from one Wikipedia to another~\cite{Stvilia2009IQEval}. If we change our basis to geographic areas, we find that certain areas (e.g a country like India) can span multiple communities and thereby enable an analysis of cultural factors behind those communities. In their 2022 paper, Khatri et al.~\cite{khatri2022social} study three Wikipedia communities in India and discover how experience with local governments and Free/Open Source Software helped one community flourish and how monetary effects impaired two other communities.

We used additional datasets to help us understand how user motivation, goals, and experience can help explain our findings. However, we left an opportunity to dig deeper into these datasets. The Welcome Survey is deployed across all Wikipedia wikis, meaning it is a rich source of data that warrants further study. We are also planning to use geographic data, for example, to learn if newcomers from certain areas are more or less likely to be successful. While some Wikipedia editions are written in geographically limited languages (e.g. the three languages of India studied by Khatri et al. mentioned above) and therefore can be said to be located in a specific region, some cannot (e.g. English and Spanish~\cite{urwikipedia2012}). Continuing this line of work of combining datasets to unlock new insights into peer production communities is important in future research.

Participating in a peer production community can mean many things. The Newcomer Homepage and Newcomer Tasks were motivated by the New Editors Experience research and Lave and Wenger's framework, meaning editing articles would be the core activity. There are many other ways to participate, such as organizing events (both online and offline), writing documentation, writing code, or helping newcomers by responding to their questions when they need help. Similarly, studies and interventions in peer production communities can also benefit from taking a holistic view of what it means to contribute.

\section{Conclusion}
In this paper, we introduced the Newcomer Homepage and Newcomer Tasks and showed how these interventions could affect newcomer participation in peer production communities through a large-scale controlled experiment. We highlighted areas where the features did not perform as expected and described changes to counteract those. Lastly, we discussed impacts and opportunities for future work, some already in motion.

While our analysis found results that were not solely positive, the Newcomer Homepage provides access to features that several threads of research deemed essential, such as making it easy to find opportunities to contribute and having access to a mentor. The features have also been continuously developed to fix the issues covered in this paper and make improvements based on community feedback. As a result, it is now the default newcomer experience on all Wikipedia wikis.\footnote{\url{https://meta.wikimedia.org/wiki/List_of_Wikipedias}}

\begin{acks}
We would first like to thank the other members of the Wikimedia Foundation's Growth team, both current and past, for the work done in designing and developing the features used in this study. In particular, we thank our community relation specialist Benoît Evellin and the partner wiki ambassadors Habib Mhenni, Martin Urbanec, User:Bluetpp, and User:-revi for their work in helping facilitate multiple experiments. We would also like to thank the community members of the 27 Wikipedia wikis that participated in the experiment used in this study and all members of the Wikimedia movement for their tireless contributions to the wikis. Lastly, we thank our CSCW reviewers for their work and excellent feedback that significantly strengthened our paper.
\end{acks}

\bibliographystyle{ACM-Reference-Format}
\bibliography{bibliography}


\begin{thebibliography}{65}


\ifx \showCODEN    \undefined \def \showCODEN     #1{\unskip}     \fi
\ifx \showDOI      \undefined \def \showDOI       #1{#1}\fi
\ifx \showISBNx    \undefined \def \showISBNx     #1{\unskip}     \fi
\ifx \showISBNxiii \undefined \def \showISBNxiii  #1{\unskip}     \fi
\ifx \showISSN     \undefined \def \showISSN      #1{\unskip}     \fi
\ifx \showLCCN     \undefined \def \showLCCN      #1{\unskip}     \fi
\ifx \shownote     \undefined \def \shownote      #1{#1}          \fi
\ifx \showarticletitle \undefined \def \showarticletitle #1{#1}   \fi
\ifx \showURL      \undefined \def \showURL       {\relax}        \fi
\providecommand\bibfield[2]{#2}
\providecommand\bibinfo[2]{#2}
\providecommand\natexlab[1]{#1}
\providecommand\showeprint[2][]{arXiv:#2}

\bibitem[Angrist and Pischke(2009)]%
        {angrist2009harmless}
\bibfield{author}{\bibinfo{person}{Joshua~D Angrist} {and}
  \bibinfo{person}{J{\"o}rn-Steffen Pischke}.} \bibinfo{year}{2009}\natexlab{}.
\newblock \bibinfo{booktitle}{\emph{Mostly Harmless Econometrics: An
  Empiricist's Companion}}.
\newblock \bibinfo{publisher}{Princeton University Press}.
\newblock


\bibitem[Anthony et~al\mbox{.}(2009)]%
        {anthony2009reputation}
\bibfield{author}{\bibinfo{person}{Denise Anthony}, \bibinfo{person}{Sean~W
  Smith}, {and} \bibinfo{person}{Timothy Williamson}.}
  \bibinfo{year}{2009}\natexlab{}.
\newblock \showarticletitle{Reputation and Reliability in Collective Goods: The
  Case of the Online Encyclopedia Wikipedia}.
\newblock \bibinfo{journal}{\emph{Rationality and Society}}
  \bibinfo{volume}{21}, \bibinfo{number}{3} (\bibinfo{year}{2009}),
  \bibinfo{pages}{283--306}.
\newblock


\bibitem[Bates et~al\mbox{.}(2015)]%
        {lme42015}
\bibfield{author}{\bibinfo{person}{Douglas Bates}, \bibinfo{person}{Martin
  M{\"a}chler}, \bibinfo{person}{Ben Bolker}, {and} \bibinfo{person}{Steve
  Walker}.} \bibinfo{year}{2015}\natexlab{}.
\newblock \showarticletitle{Fitting Linear Mixed-Effects Models Using {lme4}}.
\newblock \bibinfo{journal}{\emph{Journal of Statistical Software}}
  \bibinfo{volume}{67}, \bibinfo{number}{1} (\bibinfo{year}{2015}),
  \bibinfo{pages}{1--48}.
\newblock
\urldef\tempurl%
\url{https://doi.org/10.18637/jss.v067.i01}
\showDOI{\tempurl}


\bibitem[Benkler(2002)]%
        {benkler2002coase}
\bibfield{author}{\bibinfo{person}{Yochai Benkler}.}
  \bibinfo{year}{2002}\natexlab{}.
\newblock \showarticletitle{Coase's Penguin, or, Linux and ``the Nature of the
  Firm''}.
\newblock \bibinfo{journal}{\emph{Yale Law Journal}} (\bibinfo{year}{2002}),
  \bibinfo{pages}{369--446}.
\newblock


\bibitem[Benkler et~al\mbox{.}(2015)]%
        {benkler2015peer}
\bibfield{author}{\bibinfo{person}{Yochai Benkler}, \bibinfo{person}{Aaron
  Shaw}, {and} \bibinfo{person}{Benjamin~Mako Hill}.}
  \bibinfo{year}{2015}\natexlab{}.
\newblock \showarticletitle{Peer Production: A Form of Collective
  Intelligence}.
\newblock \bibinfo{journal}{\emph{Handbook of Collective Intelligence}}
  \bibinfo{volume}{175} (\bibinfo{year}{2015}).
\newblock


\bibitem[Brown(2015)]%
        {brown2015wpmobile}
\bibfield{author}{\bibinfo{person}{Andrew Brown}.}
  \bibinfo{year}{2015}\natexlab{}.
\newblock \bibinfo{title}{Wikipedia editors are a dying breed. The reason?
  Mobile}.
\newblock
  \bibinfo{howpublished}{\url{https://www.theguardian.com/commentisfree/2015/jun/25/wikipedia-editors-dying-breed-mobile-smartphone-technology-online-encyclopedia}}.
\newblock
\newblock
\shownote{Retrieved July 6, 2022}.


\bibitem[Bryant et~al\mbox{.}(2005)]%
        {bryant2005becoming}
\bibfield{author}{\bibinfo{person}{Susan~L Bryant}, \bibinfo{person}{Andrea
  Forte}, {and} \bibinfo{person}{Amy Bruckman}.}
  \bibinfo{year}{2005}\natexlab{}.
\newblock \showarticletitle{Becoming Wikipedian: Transformation of
  Participation in a Collaborative Online Encyclopedia}. In
  \bibinfo{booktitle}{\emph{Proceedings of GROUP}}. \bibinfo{pages}{1--10}.
\newblock


\bibitem[Burke and Kraut(2008)]%
        {burke2008mop}
\bibfield{author}{\bibinfo{person}{Moira Burke} {and} \bibinfo{person}{Robert
  Kraut}.} \bibinfo{year}{2008}\natexlab{}.
\newblock \showarticletitle{Mopping {U}p: {M}odeling {W}ikipedia {P}romotion
  {D}ecisions}. In \bibinfo{booktitle}{\emph{Proc. of CSCW}} (San Diego, CA,
  USA). \bibinfo{pages}{27--36}.
\newblock
\showISBNx{978-1-60558-007-4}
\urldef\tempurl%
\url{https://doi.org/10.1145/1460563.1460571}
\showDOI{\tempurl}


\bibitem[Bürkner(2017)]%
        {brms2017}
\bibfield{author}{\bibinfo{person}{Paul-Christian Bürkner}.}
  \bibinfo{year}{2017}\natexlab{}.
\newblock \showarticletitle{{brms}: An {R} Package for {Bayesian} Multilevel
  Models Using {Stan}}.
\newblock \bibinfo{journal}{\emph{Journal of Statistical Software}}
  \bibinfo{volume}{80}, \bibinfo{number}{1} (\bibinfo{year}{2017}),
  \bibinfo{pages}{1--28}.
\newblock
\urldef\tempurl%
\url{https://doi.org/10.18637/jss.v080.i01}
\showDOI{\tempurl}


\bibitem[Bürkner(2018)]%
        {brms2018}
\bibfield{author}{\bibinfo{person}{Paul-Christian Bürkner}.}
  \bibinfo{year}{2018}\natexlab{}.
\newblock \showarticletitle{Advanced {Bayesian} Multilevel Modeling with the
  {R} Package {brms}}.
\newblock \bibinfo{journal}{\emph{The R Journal}} \bibinfo{volume}{10},
  \bibinfo{number}{1} (\bibinfo{year}{2018}), \bibinfo{pages}{395--411}.
\newblock
\urldef\tempurl%
\url{https://doi.org/10.32614/RJ-2018-017}
\showDOI{\tempurl}


\bibitem[Chen et~al\mbox{.}(2017)]%
        {chen2017community}
\bibfield{author}{\bibinfo{person}{Chunyang Chen}, \bibinfo{person}{Zhenchang
  Xing}, {and} \bibinfo{person}{Yang Liu}.} \bibinfo{year}{2017}\natexlab{}.
\newblock \showarticletitle{By the Community \& For the Community: A Deep
  Learning Approach to Assist Collaborative Editing in Q\&A Sites}.
\newblock \bibinfo{journal}{\emph{Proceedings of the ACM on Human-Computer
  Interaction}} \bibinfo{volume}{1}, \bibinfo{number}{CSCW}
  (\bibinfo{year}{2017}), \bibinfo{pages}{1--21}.
\newblock


\bibitem[Cohen(2011)]%
        {cohen2011gender}
\bibfield{author}{\bibinfo{person}{Noam Cohen}.}
  \bibinfo{year}{2011}\natexlab{}.
\newblock \bibinfo{title}{Define Gender Gap? Look Up Wikipedia’s Contributor
  List}.
\newblock
  \bibinfo{howpublished}{\url{https://www.nytimes.com/2011/01/31/business/media/31link.html}}.
\newblock
\newblock
\shownote{Retrieved July 13, 2022}.


\bibitem[Cosley et~al\mbox{.}(2007)]%
        {cosley2007sbot}
\bibfield{author}{\bibinfo{person}{Dan Cosley}, \bibinfo{person}{Dan
  Frankowski}, \bibinfo{person}{Loren Terveen}, {and} \bibinfo{person}{John
  Riedl}.} \bibinfo{year}{2007}\natexlab{}.
\newblock \showarticletitle{Suggest{B}ot: {U}sing {I}ntelligent {T}ask
  {R}outing to {H}elp {P}eople {F}ind {W}ork in {W}ikipedia}. In
  \bibinfo{booktitle}{\emph{Proc. IUI}} (Honolulu, Hawaii, USA).
  \bibinfo{pages}{32--41}.
\newblock
\showISBNx{1-59593-481-2}
\urldef\tempurl%
\url{https://doi.org/10.1145/1216295.1216309}
\showDOI{\tempurl}


\bibitem[Dittus and Capra(2017)]%
        {dittus2017private}
\bibfield{author}{\bibinfo{person}{Martin Dittus} {and} \bibinfo{person}{Licia
  Capra}.} \bibinfo{year}{2017}\natexlab{}.
\newblock \showarticletitle{Private Peer Feedback as Engagement Driver in
  Humanitarian Mapping}.
\newblock \bibinfo{journal}{\emph{Proc. ACM Hum.-Comput. Interact.}}
  \bibinfo{volume}{1}, \bibinfo{number}{CSCW}, Article \bibinfo{articleno}{40}
  (\bibinfo{date}{dec} \bibinfo{year}{2017}), \bibinfo{numpages}{18}~pages.
\newblock
\urldef\tempurl%
\url{https://doi.org/10.1145/3134675}
\showDOI{\tempurl}


\bibitem[Dittus et~al\mbox{.}(2016)]%
        {dittus2016analysing}
\bibfield{author}{\bibinfo{person}{Martin Dittus}, \bibinfo{person}{Giovanni
  Quattrone}, {and} \bibinfo{person}{Licia Capra}.}
  \bibinfo{year}{2016}\natexlab{}.
\newblock \showarticletitle{Analysing volunteer engagement in humanitarian
  mapping: building contributor communities at large scale}. In
  \bibinfo{booktitle}{\emph{Proceedings of the 19th ACM Conference on
  Computer-Supported Cooperative Work \& Social Computing}}.
  \bibinfo{pages}{108--118}.
\newblock


\bibitem[Economist(2021)]%
        {economist2021wp20}
\bibfield{author}{\bibinfo{person}{The Economist}.}
  \bibinfo{year}{2021}\natexlab{}.
\newblock \bibinfo{title}{Wikipedia is 20, and its reputation has never been
  higher}.
\newblock
  \bibinfo{howpublished}{\url{https://www.economist.com/international/2021/01/09/wikipedia-is-20-and-its-reputation-has-never-been-higher}}.
\newblock
\newblock
\shownote{Retrieved July 6, 2022}.


\bibitem[Ford et~al\mbox{.}(2016)]%
        {ford2016paradise}
\bibfield{author}{\bibinfo{person}{Denae Ford}, \bibinfo{person}{Justin Smith},
  \bibinfo{person}{Philip~J. Guo}, {and} \bibinfo{person}{Chris Parnin}.}
  \bibinfo{year}{2016}\natexlab{}.
\newblock \showarticletitle{Paradise Unplugged: Identifying Barriers for Female
  Participation on Stack Overflow}. In \bibinfo{booktitle}{\emph{Proceedings of
  the 2016 24th ACM SIGSOFT International Symposium on Foundations of Software
  Engineering}}. \bibinfo{address}{New York, NY, USA},
  \bibinfo{numpages}{12}~pages.
\newblock
\showISBNx{9781450342186}
\urldef\tempurl%
\url{https://doi.org/10.1145/2950290.2950331}
\showDOI{\tempurl}


\bibitem[Forte et~al\mbox{.}(2017)]%
        {forte2017privacy}
\bibfield{author}{\bibinfo{person}{Andrea Forte}, \bibinfo{person}{Nazanin
  Andalibi}, {and} \bibinfo{person}{Rachel Greenstadt}.}
  \bibinfo{year}{2017}\natexlab{}.
\newblock \showarticletitle{Privacy, anonymity, and perceived risk in open
  collaboration: A study of Tor users and Wikipedians}. In
  \bibinfo{booktitle}{\emph{Proceedings of CSCW}}. \bibinfo{pages}{1800--1811}.
\newblock


\bibitem[Foundation(2014)]%
        {WMFanonresearch}
\bibfield{author}{\bibinfo{person}{Wikimedia Foundation}.}
  \bibinfo{year}{2014}\natexlab{}.
\newblock \bibinfo{title}{Research:Asking anonymous editors to register}.
\newblock
  \bibinfo{howpublished}{\url{https://meta.wikimedia.org/w/index.php?title=Research:Asking_anonymous_editors_to_register&oldid=18839498}}.
\newblock
\newblock
\shownote{Retrieved July 11, 2022}.


\bibitem[Foundation(2021)]%
        {WMFdataretention}
\bibfield{author}{\bibinfo{person}{Wikimedia Foundation}.}
  \bibinfo{year}{2021}\natexlab{}.
\newblock \bibinfo{title}{Data Retention Guidelines}.
\newblock
  \bibinfo{howpublished}{\url{https://meta.wikimedia.org/w/index.php?title=Data_retention_guidelines&oldid=21009378}}.
\newblock
\newblock
\shownote{Retrieved July 6, 2022}.


\bibitem[Fox et~al\mbox{.}(2021)]%
        {ivreg2021}
\bibfield{author}{\bibinfo{person}{John Fox}, \bibinfo{person}{Christian
  Kleiber}, \bibinfo{person}{Achim Zeileis}, {and} \bibinfo{person}{Nikolas
  Kuschnig}.} \bibinfo{year}{2021}\natexlab{}.
\newblock \bibinfo{title}{ivreg: Instrumental-Variables Regression}.
\newblock
  \bibinfo{howpublished}{\url{https://cran.r-project.org/package=ivreg}}.
\newblock
\newblock
\shownote{Retrieved January 14, 2023}.


\bibitem[Gallus(2017)]%
        {gallus2017fostering}
\bibfield{author}{\bibinfo{person}{Jana Gallus}.}
  \bibinfo{year}{2017}\natexlab{}.
\newblock \showarticletitle{Fostering Public Good Contributions with Symbolic
  Awards: A Large-Scale Natural Field Experiment at Wikipedia}.
\newblock \bibinfo{journal}{\emph{Management Science}} \bibinfo{volume}{63},
  \bibinfo{number}{12} (\bibinfo{year}{2017}), \bibinfo{pages}{3999--4015}.
\newblock


\bibitem[Geiger and Halfaker(2013)]%
        {geiger2013levee}
\bibfield{author}{\bibinfo{person}{R~Stuart Geiger} {and}
  \bibinfo{person}{Aaron Halfaker}.} \bibinfo{year}{2013}\natexlab{}.
\newblock \showarticletitle{When the Levee Breaks: Without Bots, What Happens
  to Wikipedia's Quality Control Processes?}. In
  \bibinfo{booktitle}{\emph{Proceedings of OpenSym}}.
\newblock


\bibitem[Geiger and Ribes(2010)]%
        {Geiger2010}
\bibfield{author}{\bibinfo{person}{R.~Stuart Geiger} {and}
  \bibinfo{person}{David Ribes}.} \bibinfo{year}{2010}\natexlab{}.
\newblock \showarticletitle{The work of sustaining order in {Wikipedia}: the
  banning of a vandal}. In \bibinfo{booktitle}{\emph{Proc. CSCW}}.
  \bibinfo{pages}{117--126}.
\newblock
\showISBNx{978-1-60558-795-0}
\urldef\tempurl%
\url{https://doi.org/10.1145/1718918.1718941}
\showDOI{\tempurl}


\bibitem[Gelman and Hill(2006)]%
        {gelmanhill2006}
\bibfield{author}{\bibinfo{person}{Andrew Gelman} {and}
  \bibinfo{person}{Jennifer Hill}.} \bibinfo{year}{2006}\natexlab{}.
\newblock \bibinfo{booktitle}{\emph{Data Analysis Using Regression and
  Multilevel/Hierarchical Models}}.
\newblock \bibinfo{publisher}{Cambridge University Press}.
\newblock


\bibitem[Graham et~al\mbox{.}(2014)]%
        {graham2014geography}
\bibfield{author}{\bibinfo{person}{Mark Graham}, \bibinfo{person}{Bernie
  Hogan}, \bibinfo{person}{Ralph~K Straumann}, {and} \bibinfo{person}{Ahmed
  Medhat}.} \bibinfo{year}{2014}\natexlab{}.
\newblock \showarticletitle{Uneven geographies of user-generated information:
  Patterns of increasing informational poverty}.
\newblock \bibinfo{journal}{\emph{Annals of the Association of American
  Geographers}} \bibinfo{volume}{104}, \bibinfo{number}{4}
  (\bibinfo{year}{2014}), \bibinfo{pages}{746--764}.
\newblock


\bibitem[Haklay and Weber(2008)]%
        {haklay2008osm}
\bibfield{author}{\bibinfo{person}{M. Haklay} {and} \bibinfo{person}{P.
  Weber}.} \bibinfo{year}{2008}\natexlab{}.
\newblock \showarticletitle{OpenStreetMap: User-Generated Street Maps}.
\newblock \bibinfo{journal}{\emph{IEEE Pervasive Computing}}
  \bibinfo{volume}{7}, \bibinfo{number}{4} (\bibinfo{date}{Oct}
  \bibinfo{year}{2008}), \bibinfo{pages}{12--18}.
\newblock
\showISSN{1536-1268}
\urldef\tempurl%
\url{https://doi.org/10.1109/MPRV.2008.80}
\showDOI{\tempurl}


\bibitem[Halfaker(2017)]%
        {halfaker2017keilana}
\bibfield{author}{\bibinfo{person}{Aaron Halfaker}.}
  \bibinfo{year}{2017}\natexlab{}.
\newblock \showarticletitle{Interpolating quality dynamics in Wikipedia and
  demonstrating the Keilana effect}. In \bibinfo{booktitle}{\emph{Proceedings
  of OpenSym}}.
\newblock


\bibitem[Halfaker and Geiger(2020)]%
        {halfaker2020ores}
\bibfield{author}{\bibinfo{person}{Aaron Halfaker} {and}
  \bibinfo{person}{R~Stuart Geiger}.} \bibinfo{year}{2020}\natexlab{}.
\newblock \showarticletitle{{ORES}: {L}owering {B}arriers with {P}articipatory
  {M}achine {L}earning in {W}ikipedia}.
\newblock \bibinfo{journal}{\emph{Proceedings of the ACM on Human-Computer
  Interaction}} \bibinfo{volume}{4}, \bibinfo{number}{CSCW2}
  (\bibinfo{year}{2020}), \bibinfo{pages}{1--37}.
\newblock


\bibitem[Halfaker et~al\mbox{.}(2013)]%
        {halfaker2013abs}
\bibfield{author}{\bibinfo{person}{Aaron Halfaker}, \bibinfo{person}{R.~Stuart
  Geiger}, \bibinfo{person}{Jonathan~T. Morgan}, {and} \bibinfo{person}{John
  Riedl}.} \bibinfo{year}{2013}\natexlab{}.
\newblock \showarticletitle{The {R}ise and {D}ecline of an {O}pen
  {C}ollaboration {S}ystem: {H}ow {W}ikipedia's {R}eaction to {P}opularity {I}s
  {C}ausing {I}ts {D}ecline}.
\newblock \bibinfo{journal}{\emph{American Behavioral Scientist}}
  \bibinfo{volume}{57}, \bibinfo{number}{5} (\bibinfo{year}{2013}),
  \bibinfo{pages}{664--688}.
\newblock
\urldef\tempurl%
\url{https://doi.org/10.1177/0002764212469365}
\showDOI{\tempurl}
\showeprint{http://abs.sagepub.com/content/57/5/664.full.pdf+html}


\bibitem[Halfaker et~al\mbox{.}(2009)]%
        {Halfaker2009}
\bibfield{author}{\bibinfo{person}{Aaron Halfaker}, \bibinfo{person}{Aniket
  Kittur}, \bibinfo{person}{Robert Kraut}, {and} \bibinfo{person}{John Riedl}.}
  \bibinfo{year}{2009}\natexlab{}.
\newblock \showarticletitle{A jury of your peers: quality, experience and
  ownership in {Wikipedia}}. In \bibinfo{booktitle}{\emph{Proc. WikiSym}}.
\newblock
\showISBNx{978-1-60558-730-1}
\urldef\tempurl%
\url{https://doi.org/10.1145/1641309.1641332}
\showDOI{\tempurl}


\bibitem[Hargittai and Shaw(2015)]%
        {hargittai2015gender}
\bibfield{author}{\bibinfo{person}{Eszter Hargittai} {and}
  \bibinfo{person}{Aaron Shaw}.} \bibinfo{year}{2015}\natexlab{}.
\newblock \showarticletitle{Mind the skills gap: the role of Internet know-how
  and gender in differentiated contributions to Wikipedia}.
\newblock \bibinfo{journal}{\emph{Information, Communication \& Society}}
  \bibinfo{volume}{18}, \bibinfo{number}{4} (\bibinfo{year}{2015}),
  \bibinfo{pages}{424--442}.
\newblock


\bibitem[Hecht and Gergle(2009)]%
        {HechtCT2009Bias}
\bibfield{author}{\bibinfo{person}{Brent Hecht} {and} \bibinfo{person}{Darren
  Gergle}.} \bibinfo{year}{2009}\natexlab{}.
\newblock \showarticletitle{Measuring self-focus bias in community-maintained
  knowledge repositories}. In \bibinfo{booktitle}{\emph{Proc. C\&T}}.
  \bibinfo{pages}{11--20}.
\newblock


\bibitem[Hill and Shaw(2013)]%
        {makoshaw2013gender}
\bibfield{author}{\bibinfo{person}{Benjamin~Mako Hill} {and}
  \bibinfo{person}{Aaron Shaw}.} \bibinfo{year}{2013}\natexlab{}.
\newblock \showarticletitle{{The Wikipedia Gender Gap Revisited: Characterizing
  Survey Response Bias with Propensity Score Estimation}}.
\newblock \bibinfo{journal}{\emph{PLoS ONE}} \bibinfo{volume}{8},
  \bibinfo{number}{6} (\bibinfo{date}{06} \bibinfo{year}{2013}),
  \bibinfo{pages}{1--5}.
\newblock
\urldef\tempurl%
\url{https://doi.org/10.1371/journal.pone.0065782}
\showDOI{\tempurl}


\bibitem[Hill and Shaw(2021)]%
        {hill2021hidden}
\bibfield{author}{\bibinfo{person}{Benjamin~Mako Hill} {and}
  \bibinfo{person}{Aaron Shaw}.} \bibinfo{year}{2021}\natexlab{}.
\newblock \showarticletitle{The Hidden Costs of Requiring Accounts:
  Quasi-Experimental Evidence From Peer Production}.
\newblock \bibinfo{journal}{\emph{Communication Research}}
  \bibinfo{volume}{48}, \bibinfo{number}{6} (\bibinfo{year}{2021}),
  \bibinfo{pages}{771--795}.
\newblock


\bibitem[Hristova et~al\mbox{.}(2013)]%
        {hristova2013osm}
\bibfield{author}{\bibinfo{person}{Desislava Hristova},
  \bibinfo{person}{Giovanni Quattrone}, \bibinfo{person}{Afra~J Mashhadi},
  {and} \bibinfo{person}{Licia Capra}.} \bibinfo{year}{2013}\natexlab{}.
\newblock \showarticletitle{{The Life of the Party: Impact of Social Mapping in
  OpenStreetMap}}. In \bibinfo{booktitle}{\emph{Proc. of ICWSM}}.
\newblock


\bibitem[Huang et~al\mbox{.}(2015)]%
        {huang2015activists}
\bibfield{author}{\bibinfo{person}{Shih-Wen Huang}, \bibinfo{person}{Minhyang
  Suh}, \bibinfo{person}{Benjamin~Mako Hill}, {and} \bibinfo{person}{Gary
  Hsieh}.} \bibinfo{year}{2015}\natexlab{}.
\newblock \showarticletitle{How Activists Are Both Born and Made: An Analysis
  of Users on Change.org}. In \bibinfo{booktitle}{\emph{Proceedings of CHI}}.
  \bibinfo{pages}{211--220}.
\newblock


\bibitem[Jackson et~al\mbox{.}(2018)]%
        {jackson2018did}
\bibfield{author}{\bibinfo{person}{Corey~Brian Jackson}, \bibinfo{person}{Kevin
  Crowston}, {and} \bibinfo{person}{Carsten {\O}sterlund}.}
  \bibinfo{year}{2018}\natexlab{}.
\newblock \showarticletitle{Did they login? Patterns of Anonymous Contributions
  in Online Communities}.
\newblock \bibinfo{journal}{\emph{Proceedings of the ACM on Human-Computer
  Interaction}} \bibinfo{volume}{2}, \bibinfo{number}{CSCW}
  (\bibinfo{year}{2018}), \bibinfo{pages}{1--16}.
\newblock


\bibitem[Johnson et~al\mbox{.}(2016)]%
        {johnson2016}
\bibfield{author}{\bibinfo{person}{Isaac~L. Johnson}, \bibinfo{person}{Yilun
  Lin}, \bibinfo{person}{Toby Jia-Jun Li}, \bibinfo{person}{Andrew Hall},
  \bibinfo{person}{Aaron Halfaker}, \bibinfo{person}{Johannes Sch\"{o}ning},
  {and} \bibinfo{person}{Brent Hecht}.} \bibinfo{year}{2016}\natexlab{}.
\newblock \showarticletitle{{Not at Home on the Range: Peer Production and the
  Urban/Rural Divide}}. In \bibinfo{booktitle}{\emph{Proc. of CHI}}.
  \bibinfo{pages}{13--25}.
\newblock
\showISBNx{978-1-4503-3362-7}
\urldef\tempurl%
\url{https://doi.org/10.1145/2858036.2858123}
\showDOI{\tempurl}


\bibitem[Khatri et~al\mbox{.}(2022)]%
        {khatri2022social}
\bibfield{author}{\bibinfo{person}{Sejal Khatri}, \bibinfo{person}{Aaron Shaw},
  \bibinfo{person}{Sayamindu Dasgupta}, {and} \bibinfo{person}{Benjamin~Mako
  Hill}.} \bibinfo{year}{2022}\natexlab{}.
\newblock \showarticletitle{The Social Embeddedness of Peer Production: A
  Comparative Qualitative Analysis of Three Indian Language Wikipedia
  Editions}. In \bibinfo{booktitle}{\emph{Proceedings of the 2022 CHI
  Conference on Human Factors in Computing Systems}}.
\newblock
\showISBNx{9781450391573}
\urldef\tempurl%
\url{https://doi.org/10.1145/3491102.3501832}
\showDOI{\tempurl}


\bibitem[Klein et~al\mbox{.}(2016)]%
        {klein2016whgi}
\bibfield{author}{\bibinfo{person}{Maximilian Klein}, \bibinfo{person}{Harsh
  Gupta}, \bibinfo{person}{Vivek Rai}, \bibinfo{person}{Piotr Konieczny}, {and}
  \bibinfo{person}{Haiyi Zhu}.} \bibinfo{year}{2016}\natexlab{}.
\newblock \showarticletitle{Monitoring the Gender Gap with Wikidata Human
  Gender Indicators}. In \bibinfo{booktitle}{\emph{Proceedings of the 12th
  International Symposium on Open Collaboration}}.
\newblock
\showISBNx{9781450344517}
\urldef\tempurl%
\url{https://doi.org/10.1145/2957792.2957798}
\showDOI{\tempurl}


\bibitem[Konieczny and Klein(2018)]%
        {konieczny2018gender}
\bibfield{author}{\bibinfo{person}{Piotr Konieczny} {and}
  \bibinfo{person}{Maximilian Klein}.} \bibinfo{year}{2018}\natexlab{}.
\newblock \showarticletitle{Gender gap through time and space: A journey
  through Wikipedia biographies via the Wikidata Human Gender Indicator}.
\newblock \bibinfo{journal}{\emph{New Media \& Society}} \bibinfo{volume}{20},
  \bibinfo{number}{12} (\bibinfo{year}{2018}), \bibinfo{pages}{4608--4633}.
\newblock


\bibitem[Lam et~al\mbox{.}(2011)]%
        {lam2011gender}
\bibfield{author}{\bibinfo{person}{Shyong (Tony)~K. Lam},
  \bibinfo{person}{Anuradha Uduwage}, \bibinfo{person}{Zhenhua Dong},
  \bibinfo{person}{Shilad Sen}, \bibinfo{person}{David~R. Musicant},
  \bibinfo{person}{Loren Terveen}, {and} \bibinfo{person}{John Riedl}.}
  \bibinfo{year}{2011}\natexlab{}.
\newblock \showarticletitle{{WP}:{C}lubhouse?: An {E}xploration of
  {W}ikipedia's {G}ender {I}mbalance}. In \bibinfo{booktitle}{\emph{Proc. of
  WikiSym}}. \bibinfo{pages}{1--10}.
\newblock


\bibitem[Lave and Wenger(1991)]%
        {lw1991cop}
\bibfield{author}{\bibinfo{person}{Jean Lave} {and} \bibinfo{person}{Etienne
  Wenger}.} \bibinfo{year}{1991}\natexlab{}.
\newblock \bibinfo{booktitle}{\emph{Situated learning: {Legitimate} peripheral
  participation}}.
\newblock \bibinfo{publisher}{Cambridge University Press}.
\newblock


\bibitem[Lee et~al\mbox{.}(2017)]%
        {lee2017understanding}
\bibfield{author}{\bibinfo{person}{Amanda Lee}, \bibinfo{person}{Jeffrey~C.
  Carver}, {and} \bibinfo{person}{Amiangshu Bosu}.}
  \bibinfo{year}{2017}\natexlab{}.
\newblock \showarticletitle{Understanding the Impressions, Motivations, and
  Barriers of One Time Code Contributors to FLOSS Projects: A Survey}. In
  \bibinfo{booktitle}{\emph{2017 IEEE/ACM 39th International Conference on
  Software Engineering (ICSE)}}. \bibinfo{pages}{187--197}.
\newblock
\urldef\tempurl%
\url{https://doi.org/10.1109/ICSE.2017.25}
\showDOI{\tempurl}


\bibitem[Li et~al\mbox{.}(2020)]%
        {li2020wikied}
\bibfield{author}{\bibinfo{person}{Ang Li}, \bibinfo{person}{Zheng Yao},
  \bibinfo{person}{Diyi Yang}, \bibinfo{person}{Chinmay Kulkarni},
  \bibinfo{person}{Rosta Farzan}, {and} \bibinfo{person}{Robert~E. Kraut}.}
  \bibinfo{year}{2020}\natexlab{}.
\newblock \showarticletitle{Successful Online Socialization: Lessons from the
  Wikipedia Education Program}.
\newblock \bibinfo{journal}{\emph{Proc. ACM Hum.-Comput. Interact.}}
  \bibinfo{volume}{4}, \bibinfo{number}{CSCW1}, Article \bibinfo{articleno}{50}
  (\bibinfo{year}{2020}).
\newblock
\urldef\tempurl%
\url{https://doi.org/10.1145/3392857}
\showDOI{\tempurl}


\bibitem[Mahmud et~al\mbox{.}(2022)]%
        {mahmud2022revisiting}
\bibfield{author}{\bibinfo{person}{Zarif Mahmud}, \bibinfo{person}{Aarjav
  Chauhan}, \bibinfo{person}{Dipto Sarkar}, {and} \bibinfo{person}{Robert
  Soden}.} \bibinfo{year}{2022}\natexlab{}.
\newblock \showarticletitle{Revisiting Engagement in Humanitarian Mapping: An
  Updated Analysis of Contributor Retention in OpenStreetMap}. In
  \bibinfo{booktitle}{\emph{CHI Conference on Human Factors in Computing
  Systems Extended Abstracts}}. \bibinfo{pages}{1--6}.
\newblock


\bibitem[Mendez et~al\mbox{.}(2018)]%
        {mendez2018open}
\bibfield{author}{\bibinfo{person}{Christopher Mendez},
  \bibinfo{person}{Hema~Susmita Padala}, \bibinfo{person}{Zoe Steine-Hanson},
  \bibinfo{person}{Claudia Hilderbrand}, \bibinfo{person}{Amber Horvath},
  \bibinfo{person}{Charles Hill}, \bibinfo{person}{Logan Simpson},
  \bibinfo{person}{Nupoor Patil}, \bibinfo{person}{Anita Sarma}, {and}
  \bibinfo{person}{Margaret Burnett}.} \bibinfo{year}{2018}\natexlab{}.
\newblock \showarticletitle{Open Source Barriers to Entry, Revisited: A
  Sociotechnical Perspective}. In \bibinfo{booktitle}{\emph{Proceedings of the
  40th International Conference on Software Engineering}}
  \emph{(\bibinfo{series}{ICSE '18})}.
\newblock
\showISBNx{9781450356381}
\urldef\tempurl%
\url{https://doi.org/10.1145/3180155.3180241}
\showDOI{\tempurl}


\bibitem[Menking et~al\mbox{.}(2019)]%
        {menking2019people}
\bibfield{author}{\bibinfo{person}{Amanda Menking}, \bibinfo{person}{Ingrid
  Erickson}, {and} \bibinfo{person}{Wanda Pratt}.}
  \bibinfo{year}{2019}\natexlab{}.
\newblock \showarticletitle{People Who Can Take It: How Women Wikipedians
  Negotiate and Navigate Safety}. In \bibinfo{booktitle}{\emph{Proceedings of
  CHI}}. \bibinfo{pages}{1--14}.
\newblock


\bibitem[Morgan et~al\mbox{.}(2013)]%
        {morgan2013tea}
\bibfield{author}{\bibinfo{person}{Jonathan~T. Morgan}, \bibinfo{person}{Siko
  Bouterse}, \bibinfo{person}{Heather Walls}, {and} \bibinfo{person}{Sarah
  Stierch}.} \bibinfo{year}{2013}\natexlab{}.
\newblock \showarticletitle{Tea and Sympathy: Crafting Positive New User
  Experiences on Wikipedia}. In \bibinfo{booktitle}{\emph{Proc. of CSCW}}.
  \bibinfo{pages}{839--848}.
\newblock


\bibitem[Morgan and Filippova(2018)]%
        {morgan2018welcome}
\bibfield{author}{\bibinfo{person}{Jonathan~T. Morgan} {and}
  \bibinfo{person}{Anna Filippova}.} \bibinfo{year}{2018}\natexlab{}.
\newblock \showarticletitle{'Welcome' Changes? Descriptive and Injunctive Norms
  in a Wikipedia Sub-Community}.
\newblock \bibinfo{journal}{\emph{Proc. ACM Hum.-Comput. Interact.}}
  \bibinfo{volume}{2}, \bibinfo{number}{CSCW}, Article \bibinfo{articleno}{52}
  (\bibinfo{date}{nov} \bibinfo{year}{2018}), \bibinfo{numpages}{26}~pages.
\newblock
\urldef\tempurl%
\url{https://doi.org/10.1145/3274321}
\showDOI{\tempurl}


\bibitem[Morgan and Halfaker(2018)]%
        {morgan2018teahouse}
\bibfield{author}{\bibinfo{person}{Jonathan~T. Morgan} {and}
  \bibinfo{person}{Aaron Halfaker}.} \bibinfo{year}{2018}\natexlab{}.
\newblock \showarticletitle{Evaluating the impact of the Wikipedia Teahouse on
  newcomer socialization and retention}. In
  \bibinfo{booktitle}{\emph{Proceedings of OpenSym}}. \bibinfo{pages}{1--7}.
\newblock


\bibitem[Musicant et~al\mbox{.}(2011)]%
        {musicant2011mentoring}
\bibfield{author}{\bibinfo{person}{David~R Musicant}, \bibinfo{person}{Yuqing
  Ren}, \bibinfo{person}{James~A Johnson}, {and} \bibinfo{person}{John Riedl}.}
  \bibinfo{year}{2011}\natexlab{}.
\newblock \showarticletitle{Mentoring in Wikipedia: a Clash of Cultures}. In
  \bibinfo{booktitle}{\emph{Proceedings of WikiSym}}.
  \bibinfo{pages}{173--182}.
\newblock


\bibitem[Narayan et~al\mbox{.}(2017)]%
        {narayan2017}
\bibfield{author}{\bibinfo{person}{Sneha Narayan}, \bibinfo{person}{Jake
  Orlowitz}, \bibinfo{person}{Jonathan Morgan}, \bibinfo{person}{Benjamin~Mako
  Hill}, {and} \bibinfo{person}{Aaron Shaw}.} \bibinfo{year}{2017}\natexlab{}.
\newblock \showarticletitle{The Wikipedia Adventure: Field Evaluation of an
  Interactive Tutorial for New Users}. In \bibinfo{booktitle}{\emph{Proceedings
  of CSCW}}. \bibinfo{pages}{1785--1799}.
\newblock


\bibitem[Okoli et~al\mbox{.}(2012)]%
        {okoli2012wikipedia}
\bibfield{author}{\bibinfo{person}{Chitu Okoli}, \bibinfo{person}{Mohamad
  Mehdi}, \bibinfo{person}{Mostafa Mesgari}, \bibinfo{person}{Finn~{\AA}rup
  Nielsen}, {and} \bibinfo{person}{Arto Lanam{\"a}ki}.}
  \bibinfo{year}{2012}\natexlab{}.
\newblock \showarticletitle{The people’s encyclopedia under the gaze of the
  sages: A systematic review of scholarly research on Wikipedia}.
\newblock \bibinfo{journal}{\emph{Available at SSRN 2021326}}
  (\bibinfo{year}{2012}).
\newblock


\bibitem[Okoli et~al\mbox{.}(2014)]%
        {okoli2014wikipedia}
\bibfield{author}{\bibinfo{person}{Chitu Okoli}, \bibinfo{person}{Mohamad
  Mehdi}, \bibinfo{person}{Mostafa Mesgari}, \bibinfo{person}{Finn~{\AA}rup
  Nielsen}, {and} \bibinfo{person}{Arto Lanam{\"a}ki}.}
  \bibinfo{year}{2014}\natexlab{}.
\newblock \showarticletitle{Wikipedia in the eyes of its beholders: A
  systematic review of scholarly research on Wikipedia readers and readership}.
\newblock \bibinfo{journal}{\emph{Journal of the Association for Information
  Science and Technology}} \bibinfo{volume}{65}, \bibinfo{number}{12}
  (\bibinfo{year}{2014}), \bibinfo{pages}{2381--2403}.
\newblock


\bibitem[Panciera et~al\mbox{.}(2009)]%
        {panciera2009}
\bibfield{author}{\bibinfo{person}{Katherine Panciera}, \bibinfo{person}{Aaron
  Halfaker}, {and} \bibinfo{person}{Loren Terveen}.}
  \bibinfo{year}{2009}\natexlab{}.
\newblock \showarticletitle{Wikipedians {A}re {B}orn, {N}ot {M}ade: {A} {S}tudy
  of {P}ower {E}ditors on {W}ikipedia}. In \bibinfo{booktitle}{\emph{Proc.
  GROUP}}. \bibinfo{pages}{51--60}.
\newblock


\bibitem[Reagle and Rhue(2011)]%
        {reagle2011gender}
\bibfield{author}{\bibinfo{person}{Joseph Reagle} {and} \bibinfo{person}{Lauren
  Rhue}.} \bibinfo{year}{2011}\natexlab{}.
\newblock \showarticletitle{Gender {Bias} in {Wikipedia} and {Britannica}}.
\newblock \bibinfo{journal}{\emph{International Journal of Communication}}
  \bibinfo{volume}{5}, \bibinfo{number}{0} (\bibinfo{year}{2011}).
\newblock
\showISSN{1932-8036}


\bibitem[Reboot and the Wikimedia~Foundation(2017)]%
        {WMFNEE2017}
\bibfield{author}{\bibinfo{person}{Reboot} {and} \bibinfo{person}{the
  Wikimedia~Foundation}.} \bibinfo{year}{2017}\natexlab{}.
\newblock \bibinfo{title}{New Editor Experiences}.
\newblock
  \bibinfo{howpublished}{\url{https://commons.wikimedia.org/wiki/File:New_Editor_Experiences_summary_of_findings,_August_2017.pdf}}.
\newblock
\newblock
\shownote{Retrieved July 5, 2022}.


\bibitem[Steinmacher et~al\mbox{.}(2018)]%
        {steinmacher2018flosscoach}
\bibfield{author}{\bibinfo{person}{Igor Steinmacher},
  \bibinfo{person}{Christoph Treude}, {and} \bibinfo{person}{Marco~Aurelio
  Gerosa}.} \bibinfo{year}{2018}\natexlab{}.
\newblock \showarticletitle{Let me in: Guidelines for the Successful Onboarding
  of Newcomers to Open Source Projects}.
\newblock \bibinfo{journal}{\emph{IEEE Software}} \bibinfo{volume}{36},
  \bibinfo{number}{4} (\bibinfo{year}{2018}), \bibinfo{pages}{41--49}.
\newblock


\bibitem[Stephens(2013)]%
        {stephens2013gender}
\bibfield{author}{\bibinfo{person}{Monica Stephens}.}
  \bibinfo{year}{2013}\natexlab{}.
\newblock \showarticletitle{Gender and the {GeoWeb:} divisions in the
  production of user-generated cartographic information}.
\newblock \bibinfo{journal}{\emph{{GeoJournal}}} (\bibinfo{year}{2013}),
  \bibinfo{pages}{1--16}.
\newblock
\showISSN{0343-2521, 1572-9893}
\urldef\tempurl%
\url{https://doi.org/10.1007/s10708-013-9492-z}
\showDOI{\tempurl}
\newblock
\shownote{00001}.


\bibitem[Stvilia et~al\mbox{.}(2009)]%
        {Stvilia2009IQEval}
\bibfield{author}{\bibinfo{person}{Besiki Stvilia}, \bibinfo{person}{Abdullah
  Al-Faraj}, {and} \bibinfo{person}{Yong~Jeong Yi}.}
  \bibinfo{year}{2009}\natexlab{}.
\newblock \showarticletitle{Issues of cross-contextual information quality
  evaluation--{The} case of {Arabic}, {English}, and {Korean} {Wikipedias}}.
\newblock \bibinfo{journal}{\emph{Library {\&} Information Science Research}}
  \bibinfo{volume}{31}, \bibinfo{number}{4} (\bibinfo{year}{2009}),
  \bibinfo{pages}{232--239}.
\newblock
\showISSN{0740-8188}
\urldef\tempurl%
\url{https://doi.org/10.1016/j.lisr.2009.07.005}
\showDOI{\tempurl}


\bibitem[TeBlunthuis et~al\mbox{.}(2018)]%
        {teblunthuis2018rise}
\bibfield{author}{\bibinfo{person}{Nathan TeBlunthuis}, \bibinfo{person}{Aaron
  Shaw}, {and} \bibinfo{person}{Benjamin~Mako Hill}.}
  \bibinfo{year}{2018}\natexlab{}.
\newblock \showarticletitle{Revisiting ``The Rise and Decline'' in a Population
  of Peer Production Projects}. In \bibinfo{booktitle}{\emph{Proceedings of
  CHI}}. \bibinfo{pages}{1--7}.
\newblock


\bibitem[Wagner et~al\mbox{.}(2021)]%
        {wagner2021gender}
\bibfield{author}{\bibinfo{person}{Claudia Wagner}, \bibinfo{person}{David
  Garcia}, \bibinfo{person}{Mohsen Jadidi}, {and} \bibinfo{person}{Markus
  Strohmaier}.} \bibinfo{year}{2021}\natexlab{}.
\newblock \showarticletitle{It’s a Man’s Wikipedia? Assessing Gender
  Inequality in an Online Encyclopedia}.
\newblock \bibinfo{journal}{\emph{Proceedings of the International AAAI
  Conference on Web and Social Media}} \bibinfo{volume}{9}, \bibinfo{number}{1}
  (\bibinfo{date}{Aug.} \bibinfo{year}{2021}), \bibinfo{pages}{454--463}.
\newblock
\urldef\tempurl%
\url{https://ojs.aaai.org/index.php/ICWSM/article/view/14628}
\showURL{%
\tempurl}


\bibitem[Warncke-Wang et~al\mbox{.}(2012)]%
        {urwikipedia2012}
\bibfield{author}{\bibinfo{person}{Morten Warncke-Wang},
  \bibinfo{person}{Anuradha Uduwage}, \bibinfo{person}{Zhenhua Dong}, {and}
  \bibinfo{person}{John Riedl}.} \bibinfo{year}{2012}\natexlab{}.
\newblock \showarticletitle{In Search of the Ur-Wikipedia: Universality,
  Similarity, and Translation in the Wikipedia Inter-language Link Network}. In
  \bibinfo{booktitle}{\emph{Proceedings of WikiSym}}.
\newblock


\end{thebibliography}

\appendix
\section{Experiment Dataset}
\label{app:dataset}

The dataset used in our evaluation of the Newcomer Homepage with the Newcomer Tasks module contains data from 27 Wikipedia editions spanning February through April 2021. Table~\ref{tab:dataset} below provides an overview of the wikis included in the dataset. It lists when the feature was deployed to a given wiki, what months of the experiment registrations were included, and how many users from a given wiki are in the dataset.

During the experiment these features were also deployed to two wikis that are not included in the dataset: Basque and Esperanto Wikipedia. We exclude Basque Wikipedia because 100\% of new registrations get the features there, meaning it has no control group. Deployment to Esperanto Wikipedia happened on 2021-03-24 in a group with Albanian, Hindi, Japanese, Norwegian bokmål, and Telugu Wikipedia. We excluded Esperanto Wikipedia because there were only a handful of registrations on that wiki in April 2021.

All wikis in this dataset randomly assigned new registrations to the treatment and control groups with an 80\% probability of treatment group assignment, as discussed in Section~\ref{subsec:datasets_metrics_methods}.

\begin{table}
    \centering
    \caption{Overview of Newcomer Tasks Experiment Dataset}
    \begin{tabular}{l l l r}
        \toprule
         \textbf{Language} & \textbf{Deployment date} & \textbf{Months included} & \textbf{Number of Users} \\
         \midrule
         Albanian & 2021-03-24 & Apr & 137 \\
         Arabic & 2019-11-21 & Feb--Apr & 40,491 \\
         Armenian & 2020-04-08 & Feb--Apr & 678 \\
         Bangla & 2021-02-10 & Mar--Apr & 4,231 \\
         Czech & 2019-11-21 & Feb--Apr & 4,652 \\
         Croatian & 2021-03-22 & Apr & 274 \\
         Danish & 2021-03-03 & Apr & 521 \\
         French & 2020-05-19 & Feb--Apr & 43,421 \\
         Hebrew & 2020-08-12 & Feb--Apr & 4,896 \\
         Hindi & 2021-03-24 & Apr & 918 \\
         Hungarian & 2020-04-08 & Feb--Apr & 2,993 \\
         Indonesian & 2021-03-17 & Apr & 3,511 \\
         Japanese & 2021-03-24 & Apr & 6,025 \\
         Korean & 2019-11-21 & Feb--Apr & 5,318 \\
         Norwegian bokmål & 2021-03-24 & Apr & 657 \\
         Persian & 2020-08-03 & Feb--Apr & 17,956 \\
         Polish & 2020-09-21 & Feb--Apr & 8,343 \\
         Portuguese & 2020-09-21 & Feb--Apr & 34,061 \\
         Romanian & 2021-03-03 & Apr & 869 \\
         Russian & 2020-08-27 & Feb--Apr & 30,253 \\
         Serbian & 2020-04-08 & Feb--Apr & 1,715 \\
         Swedish & 2020-09-30 & Feb--Apr & 4,137 \\
         Telugu & 2021-03-24 & Apr & 106 \\
         Thai & 2021-03-10 & Apr & 564 \\
         Turkish & 2020-10-19 & Feb--Apr & 14,060 \\
         Ukranian & 2020-04-08 & Feb--Apr & 5,041 \\
         Vietnamese & 2019-11-21 & Feb--Apr & 8,232 \\
         \midrule
         \textit{Total} & & & 244,060 \\
         \bottomrule
    \end{tabular}
    \label{tab:dataset}
\end{table}

\received{July 2022}
\received[revised]{January 2023}
\received[accepted]{March 2023}

\end{document}